\global\def\draftcontrol{0}

   \def\versionno{ normal functions }
\catcode`\@=11

\expandafter\ifx\csname draftcontrol\endcsname\relax\global\def\draftcontrol{0} 
\fi 

{\count255=\time\divide\count255 by 60 
\xdef\hourmin{\number\count255} 
\multiply\count255 by-60\advance\count255 by\time 
\xdef\hourmin{\hourmin:\ifnum\count255<10 0\fi\the\count255}} 
\def\draftdate{\number\month/\number\day/\number\year\ \ \ \hourmin } 

\newcommand\makepapertitle{\par

  \begingroup 
    \renewcommand\thefootnote{\@fnsymbol\c@footnote}% 
    \def\@makefnmark{\rlap{\@textsuperscript{\normalfont\@thefnmark}}}% 
    \long\def\@makefntext##1{\parindent 1em\noindent 
            \hb@xt@1.8em{% 
                \hss\@textsuperscript{\normalfont\@thefnmark}}##1}% 
     \newpage 
     \global\@topnum\z@   % Prevents figures from going at top of page. 
     \@makepapertitle 
     \thispagestyle{empty}\@thanks 
  \endgroup 
  \setcounter{footnote}{0}% 
  \global\let\thanks\relax 
  \global\let\makepapertitle\relax 
  \global\let\@makepapertitle\relax 
  \global\let\@thanks\@empty 
  \global\let\@author\@empty 
  \global\let\@date\@empty 
  \global\let\@title\@empty 
  \global\let\title\relax 
  \global\let\author\relax 
  \global\let\date\relax 
  \global\let\and\relax 
  \def\version{\let\version\@version\@gobble} 
} 
\def\@makepapertitle{% 
  \newpage 
   \ifnum\draftcontrol=1 {} 
   \version\versionno 
   \vskip 5em% 
   \else 
   \hfill\hbox to 3cm {\parbox{4cm}{\@pubnum}\hss}% 
   \vskip 5em% 
   \fi 
   \begin{center}% 
   \let \footnote \thanks 
      {\hskip -0\textwidth \hbox to 1\textwidth% 
        {\centerline{\Large\bf{\noindent\@title}}}}% 
     \vskip 2em% 
     {\normalsize%\large 
       \lineskip .5em% 
       \begin{tabular}[t]{c}% 
         \@author 
       \end{tabular}\par}% 
     \vskip 1.5em% 
     {\@bstract}% 
     \end{center}% 
     \vfill
     \@date%
     \vskip 1.5em%
   \par 
} 

\gdef\@pubnum{} 
\def\pubnum#1{% 
  \gdef\@pubnum{#1}} 

\gdef\@bstract{} 
\def\Abstract#1{% 
  \gdef\@bstract{% 
   \parbox{\textwidth-0pc}{% 
   \centerline{\bf Abstract}\penalty1000 
   \noindent%\abstractfont \baselineskip=12pt 
   \renewcommand\baselinestretch{1.0} 
   {#1}}} 
} 

\gdef\@email{}
\def\email#1{%
   \gdef\@email{%
   Email: {\tt #1}}
}

\def\ps@paper{\let\@mkboth\@gobbletwo% 
     \ifnum\draftcontrol=1 
        \def\@oddfoot{\hbox to \textwidth{\tiny \versionno \hfil\tiny\draftdate}% 
        \hskip -\textwidth \hbox to \textwidth{\hfil\rm\thepage\hfil}}% 
     \else\def\@oddfoot{\hbox to \textwidth{\hfil\rm\thepage\hfil}} 
     \fi 
     \let\@evenfoot\@oddfoot 
} 
\def\body{\clearpage 
          \pagestyle{paper} 
        } 
\newenvironment{acknowledgments}{% 
\vskip 1.25ex 
\addcontentsline{toc}{section}{Acknowledgments}
\noindent {\bf Acknowledgments} 
} 

\def\@version#1{\ifnum\draftcontrol=1 
\typeout{}\typeout{#1}\typeout{} 
\vskip3mm\centerline{\hbox{\fbox{\normalsize{\tt DRAFT -- #1 -- } 
                   {\draftdate}}}}\vskip3mm 
\fi} 
\let\version\@version 
\long\def\eqlabel#1{\ifnum\draftcontrol=1 
                    \tag@false  % there are some problems with multline without this 
                    \tag*{(\theequation) \hbox to -0.2cm{\hspace{0cm}\small{#1}\hss}} 
                    \refstepcounter{equation}  
                    \edef\@currentlabel{\theequation} 
                    \ltx@label{#1}          % use old LaTeX \label instead of new definition 
                    \else 
                    \label{#1} 
                    \fi 
                    } 
\let\st@bibitem\@bibitem 
\let\st@lbibitem\@lbibitem 
\ifnum\draftcontrol=1 
  \def\@bibitem#1{% 
    \st@bibitem{#1}\a@@label{#1}\ignorespaces} 
  \def\@lbibitem[#1]#2{% 
    \st@lbibitem[#1]{#2}\a@@label{#2}\ignorespaces} 
  \def\a@@label#1{% 
    \gdef\a@lab{\smash{\normalfont\small#1}} 
    \ifvmode 
      \if@inlabel 
        \global\setbox\@labels\hbox{% 
          \llap{\a@lab\let\a@lab\relax 
                \kern\@totalleftmargin\kern\marginparsep}% 
          \box\@labels}% 
      \fi 
    \fi} 
\fi 
\documentclass[12pt,letterpaper]{article} 
\usepackage[ps,matrix,arrow,frame,import,curve,color]{xy}
\usepackage{amsmath,bm,amsfonts,amssymb,array,calc,amsthm,rotating}
\usepackage[nosort]{cite} 
\usepackage{graphicx}
\usepackage{color}
\usepackage{slashbox}
\renewcommand\baselinestretch{1.25} 
\renewcommand\section{\@startsection {section}{1}{\z@}% 
                                   {-3.5ex \@plus -1ex \@minus -.2ex}% 
                                   {2.3ex \@plus.2ex}% 
                                   {\normalfont\large\bfseries}} 
\renewcommand\subsection{\@startsection{subsection}{2}{\z@}% 
                                   {-3.25ex\@plus -1ex \@minus -.2ex}% 
                                   {1.5ex \@plus .2ex}% 
                                   {\normalfont\normalsize\bfseries}} 
\renewcommand\subsubsection{\@startsection{subsubsection}{3}{\z@}% 
                                   {-3.25ex\@plus -1ex \@minus -.2ex}% 
                                   {1.5ex \@plus .2ex}% 
                                   {\normalfont\normalsize\it}} 
\renewcommand\paragraph{\@startsection{paragraph}{4}{\z@}% 
                                   {-3.25ex\@plus -1ex \@minus -.2ex}% 
                                   {1.5ex \@plus .2ex}% 
                                   {\normalfont\normalsize\bf}} 
\renewcommand\subparagraph{\@startsection{subparagraph}{5}{\z@}% 
                                   {-1.25ex\@plus -1ex \@minus -.2ex}% 
                                   {0ex \@plus .2ex}% 
                                   {\normalfont\normalsize\it}}

\numberwithin{equation}{section}

\long\def\@makecaption#1#2{%
  \vskip\abovecaptionskip
  \sbox\@tempboxa{{\bf #1:} #2}%
  \ifdim \wd\@tempboxa >\hsize
    {\small\bf #1:} {\small #2}\par
  \else
    \global \@minipagefalse
    \hb@xt@\hsize{\hfil\box\@tempboxa\hfil}%
  \fi
  \vskip\belowcaptionskip}
\renewcommand*\l@section[2]{%
  \ifnum \c@tocdepth >\z@
    \addpenalty\@secpenalty
    \addvspace{.1em \@plus\p@}%
    \setlength\@tempdima{1.5em}%
    \begingroup
      \parindent \z@ \rightskip \@pnumwidth
      \parfillskip -\@pnumwidth
      \leavevmode \bfseries
      \advance\leftskip\@tempdima
      \hskip -\leftskip
      #1\nobreak\hfil \nobreak\hb@xt@\@pnumwidth{\hss #2}\par
    \endgroup
  \fi}
\renewcommand*\l@subsection{\addvspace{-.1em \@plus\p@}\@dottedtocline{2}{1.5em}{2.3em}}
\renewcommand*\l@subsubsection{\addvspace{-.2em \@plus\p@}\@dottedtocline{3}{3.8em}{3.2em}}

\definecolor{refcol}{rgb}{0.2,0.2,0.8}
\definecolor{eqcol}{rgb}{.6,0,0}
\definecolor{purple}{cmyk}{0,1,0,0}

\gdef\@citecolor{refcol}
\gdef\@linkcolor{eqcol}
\def\colorlinkspurple{\gdef\@urlcolor{purple}}
\def\colorlinksblue{\gdef\@urlcolor{blue}}
\def\colorlinksred{\gdef\@urlcolor{red}}

\def\ie{{\it i.e.}} 
\def\eg{{\it e.g.}} 
\def\etc{{\it etc.}}

\def\revise#1       {\raisebox{-0em}{\rule{3pt}{1em}}% 
                     \marginpar{\raisebox{.5em}{\vrule width3pt\ 
                     \vrule width0pt height 0pt depth0.5em 
                     \hbox to 0cm{\hspace{0cm}{% 
                     \parbox[t]{4em}{\raggedright\footnotesize{#1}}}\hss}}}}

\def\cala         {{\cal A}}

\def\calc         {{\cal C}}

\def\calh         {{\cal H}} 
 
\def\calj         {{\cal J}} 
 
\def\call         {{\cal L}} 
 
\def\caln         {{\cal N}} 
\def\calo         {{\cal O}}

\def\calt         {{\cal T}}

\def\calw         {{\cal W}} 
\def\caly         {{\cal Y}}
\def\calz         {{\cal Z}}

\def\complex      {{\mathbb C}} 
 
\def\projective   {{\mathbb P}} 
 
\def\reals        {{\mathbb R}} 
\def\zet          {{\mathbb Z}} 

\def\del          {\partial} 
\def\delbar       {\bar\partial} 
\def\ee           {{\it e}} 
\def\ii           {{\it i}} 
 
\def\tr           {{\rm Tr}} 
 
\def\Im           {{\rm Im\hskip0.1em}}

\def\sqr#1#2{{\vcenter{\vbox{\hrule height.#2pt   
 \hbox{\vrule width.#2pt height#1pt \kern#1pt 
 \vrule width.#2pt}\hrule height.#2pt}}}}

\def\Hom{{\rm Hom}}

\def\Im{{\rm Im}}

\def\str{{\mathop{\rm Str}}}
\def\eps{\epsilon}

\def\MF{{\rm MF}}    % graded and orbifolded category

\def\fuk{{\rm Fuk}}
\def\om{\omega}

\def\Ipp{\mathord{\mathchar "0271 \kern-4.5pt \mathchar"0271}}

\def\End{\mathop{\rm End}}

\def\AJ{\mathop{\it AJ}}
\def\CH{\mathop{\rm CH}}

\catcode`\@=12 

\begin{document} 

%%% 
%%%%%% text starts here 
%%%%%%%%% 

\title{D-branes and Normal Functions}

\pubnum{%
arXiv:0709.4028}
\date{September 2007}

\author{
David R.~Morrison$^{a,b}$ and Johannes Walcher$^{c}$ \ \\[0.4cm]
\it $^a$ Center for Geometry and Theoretical Physics, Duke University, \\
\it Durham, NC 27708, USA\\[0.2cm]
\it $^b$ Departments of Mathematics and Physics, University of California,\\
\it Santa Barbara, CA 93106, USA \\[0.2cm]
\it $^c$ School of Natural Sciences, Institute for Advanced Study,\\
\it Princeton, NJ 08540, USA}

\Abstract{
We explain the B-model origin of extended Picard--Fuchs equations
satisfied by the D-brane superpotential on compact Calabi--Yau threefolds. 
Via the Abel--Jacobi map, the domainwall tension is identified with 
a Poincar\'e normal function---a transversal holomorphic section of 
the Griffiths intermediate Jacobian. Within this formalism, we derive 
the extended Picard--Fuchs equation associated with the mirror of 
the real quintic.
}
%\enlargethispage{1.5cm}

\makepapertitle

\body

\version\versionno

%\vskip 1em

\tableofcontents
%\newpage

\section{Introduction}

Mirror symmetry is a powerful tool to manipulate physical and 
mathematical data associated with Calabi--Yau manifolds. Soon after the
earliest examples of mirror symmetry \cite{CLS,GP,Aspinwall:1990xe},
a computation of the special geometry and the enumeration of rational 
curves on the quintic was made by Candelas, de la Ossa, Green and Parkes 
\cite{cdgp}. The computation was explained Hodge theoretically in 
\cite{guide}, and the verification of the enumerative predictions 
was completed in \cite{complete1,complete2}. A physics derivation 
of mirror symmetry from the worldsheet point of view has also been 
given \cite{Morrison:1995yh, hova}.

Meanwhile, D-branes have entered mirror symmetry in a variety of ways.
To name the most important, Witten showed that for open topological 
strings, cubic string field theory reduces to ordinary or holomorphic 
Chern--Simons theory \cite{wcs}. Kontsevich proposed to understand mirror 
symmetry as an equivalence of $A_\infty$-categories \cite{hms}, whose objects 
were later identified as D-branes. Strominger, Yau and 
Zaslow used D-branes to develop the geometric picture of mirror 
symmetry as a duality of torus fibrations \cite{syz}. 
Vafa and various collaborators (beginning with Gopakumar)
have shown that BPS states of 
D-branes are extremely useful invariants which carry a lot of physical 
and enumerative information \cite{gova}. Douglas has complemented the 
picture by a general formulation of stability conditions on D-brane 
categories \cite{douglas} (see also \cite{Aspinwall:2001pu}).

In the course of these developments, the established theory underlying
closed string mirror symmetry for Calabi--Yau manifolds---special 
geometry and Gromov--Witten invariants---has played a very useful
supporting role. It has, however, not always been clear whether 
D-branes would ultimately be part of the traditional picture or
how one would derive the closed string story, \eg, from D-brane 
categories. (This problem was posed already in \cite{hms}; for 
some recent work see \cite{Kapustin:2004df,MR2298823,caldararu-willerton}.)
As a physicist, one feels that in some sense, the underlying 
reason is that $A_\infty$-categories are too big. Since D-brane 
categories are defined off-shell, they carry a lot of redundant, 
gauge-dependent information. With some hindsight, one is led to ask 
the natural question: What is the invariant physical information stored 
in the derived category? 

In this paper, we give answers to these questions by picking up the Hodge 
theoretic considerations. Our main motivation is the recent realization
that at least in some cases, there is indeed invariant information 
in the open string sector beyond its cohomology. In ref.\ \cite{open},
it was shown that for a certain D-brane configuration on the quintic,%
\footnote{Very similar results appear to hold for many other 
one-parameter models \cite{private1, private2}.} the on-shell value of the 
superpotential, as a function over closed string moduli space, satisfies 
a differential equation which is an extension of the Picard--Fuchs equation 
which governs closed string mirror symmetry. According to general 
principles, this superpotential makes enumerative predictions in the 
A-model, which were subsequently verified rigorously in \cite{psw}. 
In this work, we will explain the B-model origin of this extended
Picard--Fuchs equation. Previous studies of analogous problems in local 
Calabi--Yau manifolds include \cite{agva,akv}, whose enumerative predictions 
were verified in \cite{grza,mayr}, and whose differential equations were 
discussed in \cite{lmw1, lmw2} 
(see also \cite{Govindarajan:2001zk,MR2282975}). 

The main idea to derive Picard--Fuchs equations in the context of open
strings has been implicit in many previous works. Consider for simplicity 
the case when we are wrapping a D5-brane on a curve in some second homology 
class of our Calabi--Yau manifold. Assume that this class has two isolated 
holomorphic representatives $C_+$ and $C_-$. Choose a three-chain
$\Gamma$, $\del\Gamma=C_+-C_-$ connecting those two representatives.
$C_+$ and $C_-$ correspond physically to two supersymmetric vacua of
an $\caln=1$ supersymmetric theory on the brane worldvolume. The 
tension of a BPS domainwall between the two vacua is, $\calt=\calw_+-
\calw_-$, equal to the superpotential difference, and given by the
geometric formula \cite{wittenqcd}
\begin{equation}
\eqlabel{asin}
\calw_+(z)-\calw_-(z) = \calt(z) = \int_\Gamma \Omega (z) ,
\end{equation}
where $\Omega(z)$ is the holomorphic three-form as a function of
complex structure moduli.

The Picard--Fuchs equation, $\call\Pi(z)=0$, is the (in general, system of
partial) differential equation satisfied by any period $\Pi(z)=
\int_{\Gamma^c}\Omega(z)$ of the holomorphic three-form over a 
closed three-cycle, $\del\Gamma^c=0$. When applying the Picard--Fuchs 
operator to a chain integral as in \eqref{asin}, we will not in general 
get zero. One type of non-vanishing contribution arises as a boundary 
term, but there are in general also other terms from differentiating 
the chain $\Gamma$. The {\it inhomogeneous Picard--Fuchs equation}
associated with $C_+-C_-$ is then
\begin{equation}
\eqlabel{similar}
\call\calt(z) = f(z) .
\end{equation}
The general existence of inhomogeneous Picard--Fuchs equations similar to
\eqref{similar} has been known in the mathematical literature at least as 
early as \cite{griffithsdiff}. (In dimension 1, of course, such notions
are completely classical.) A fairly recent reference with examples worked
out in dimension 2 (\ie, for K3 surfaces) is \cite{muellerdelangel}. The 
main result of the present work is a complete and mathematically rigorous 
derivation of the inhomogeneous Picard--Fuchs equation satisfied by $\calt(z)$ 
for the B-brane mirror to the real quintic. This is to our knowledge the first 
explicit example of an inhomogeneous Picard--Fuchs equation in dimension 
bigger than 2.

The particular form of the inhomogeneous Picard--Fuchs equation for
the real quintic was originally guessed in \cite{open} based on very 
restrictive monodromy properties that its solution should possess. 
Combined with the results of \cite{psw}, our derivation puts open 
string mirror symmetry for the real quintic at an equal level with 
the classical mirror theorems on rational curves in Calabi--Yau threefolds.

Before doing the computation in section \ref{main} (some details having
been deferred to the appendix), we will describe 
in section \ref{theory} how normal functions and the variation of mixed 
Hodge structure capture certain invariant information of the open string 
sector. We will not attempt a detailed comparison with the local toric 
case \cite{lmw1,lmw2}. It would be very interesting to understand better 
the relation between those works and ours, especially in regard to open 
string moduli. We also note that the insights into the relation between 
D-branes and normal functions have proven central in the recent computation 
of loop amplitudes in the open topological string using the extended 
holomorphic anomaly equation, see \cite{extended}.

In section \ref{realquintic}, we review in a self-contained manner the 
geometry of the real quintic and its mirror. This will help explain
some of the original background that led to the extended Picard--Fuchs 
equation. Alternatively, one can view our results in 
this paper as further evidence for the conjectural relation between 
the real quintic(s) and certain objects in the derived category of the
mirror quintic. This could be a starting point for establishing 
homological mirror symmetry for the quintic. We present our 
conclusions in section \ref{summary}.

\section{Normal Functions and D-branes}
\label{theory}

The urge to understand the differential equation of \cite{open} in
Hodge theoretic terms is very natural. In hindsight, it is not even
surprising that the correct framework is the theory of Poincar\'e
normal functions, applied to Calabi--Yau threefolds. That theory was 
developed by Griffiths \cite{griffiths1,griffithsdiff} as integral 
part of Hodge theory in higher dimension. Picard--Fuchs equations 
play an important role in the variation of Hodge structure and have 
been central to mirror symmetry for closed strings. So one should 
naturally have wondered about the use of normal functions in this 
context.

On the other hand, there are very good reasons to believe that 
normal functions will not be the full story for open string 
mirror symmetry computations. As is now well-accepted, D-branes
on Calabi--Yau manifolds can only be fully understood in some
sophisticated categorical framework. The D-brane superpotential,
which is the physical observable governed by the differential
equation, is realized mathematically in a fairly complicated 
way in the framework of $A_\infty$ categories \cite{lazaroiu,tomasiello}.
From this point of view, the relevance of classical Hodge theory is
not immediate at all.

The purpose of this section is to compile the main definitions and
theorems pertaining to normal functions, as well as to explain to the
best of our present understanding, the relation to the D-brane 
superpotential. We point out that our main computation in section 
\ref{main} takes this general theory as useful background, but 
does not strictly speaking depend on it.

\subsection{Normal functions attached to algebraic cycles}

For more details on normal functions, we recommend Griffiths' 
original papers \cite{griffiths1,griffithsdiff}, as well as the books
\cite{greencime,voisinbook2}. For an introduction to Hodge theory,
see \cite{voisinbook1}.

Let $(H_\zet^{2k-1},F^*H^{2k-1}_\complex)$ be an {\it integral Hodge structure} 
of odd weight $2k-1$. The {\it Griffiths intermediate Jacobian} is the complex
torus
\begin{equation}
\eqlabel{interjac}
J^{2k-1} = \frac{H^{2k-1}_\complex}{F^k H^{2k-1}_\complex\oplus H^{2k-1}_\zet} .
\end{equation}
As real torus, $J^{2k-1}$ is isomorphic to $H^{2k-1}_\reals/H^{2k-1}_\zet$, and
the complex structure on $J^{2k-1}$ arises from the identification 
$H^{2k-1}_{\complex}/F^k H^{2k-1}_\complex\cong H^{2k-1}_\reals$ as real 
vector spaces. Now if $(H_\zet^{2k-1},F^*\calh^{2k-1})$ 
is an integral {\it variation} of Hodge structure of weight $2k-1$ over some base $M$, 
we can consider a relative version of \eqref{interjac},
\begin{equation}
\eqlabel{interjacfib}
\calj^{2k-1} = \frac{\calh^{2k-1}}{F^k\calh^{2k-1}\oplus H_\zet^{2k-1}} .
\end{equation}
$\calj^{2k-1}\to M$ is known as the Griffiths intermediate
Jacobian fibration of the integral variation of Hodge structure.

A {\it normal function} of the variation of Hodge structure is a 
holomorphic section $\nu$ of the intermediate Jacobian fibration 
\eqref{interjacfib} satisfying Griffiths transversality for normal
functions\footnote{We are here omitting the regularity conditions
on normal functions that are required when the variation of Hodge
structure degenerates. Those will play only a minor role in our application.}
\begin{equation}
\eqlabel{normtrans}
\nabla\tilde\nu \in F^{k-1}\calh^{2k-1} \otimes\Omega_M ,
\end{equation}
where $\tilde\nu$ is any lift of $\nu$ to $\calh^{2k-1}$. Also, $\nabla$ 
is the Gauss--Manin connection and $\Omega_M$ the sheaf of differentials 
on $M$. It is easy to see that the condition \eqref{normtrans} is independent
of the lift. For if $\tilde\nu'$ is another lift, then 
\begin{equation}
\tilde\nu'=\tilde\nu +\eta_F+\eta_\zet ,
\end{equation}
where $\eta_F$ is a section of $F^k\calh^{2k-1}$ and $\eta_\zet$ is a
section of $H_\zet^{2k-1}$. The claim follows since $\nabla\eta_\zet=0$ and
$\nabla\eta_F\in F^{k-1}\calh^{2k-1}\otimes \Omega_M$ by Griffiths transversality
applied to $\calh^{2k-1}$.

The variation of Hodge structure of interest in this paper arises from the
deformation of complex structure of a family $\caly\to M$ 
of Calabi--Yau threefolds with typical fiber $Y$.
The interesting values of $k$ in this case are $k=1,2,$
and $3$. The intermediate Jacobians for $k=1$, $J^1$ and for $k=3$,
$J^5$ are known as the Picard variety and the Albanese variety respectively.
But when $Y$ is simply connected, $J^1=J^5=0$, and the only interesting
value is $k=2$, $2k-1=3$.

In the geometric situation, let's say in dimension $n$, not necessarily
equal to $3$, a useful source of normal functions are the {\it homologically 
trivial algebraic cycles}. Let $\calc\in \calz^k(\caly)$ be 
a relative algebraic cycle of codimension $k$, flat over $M$, \ie, 
$\calc=\sum n_i\calc^i$ is a finite integral linear combination 
of algebraic subsets of $\caly$. This cycle is ``homologically trivial'',
denoted $\calc\in\calz^k(\caly)_{\rm hom}$ if the image of $C_m=\calc\cap
 Y_m$ in $H^{2k}(Y_m;\zet)$ is trivial for all $m\in M$. (Here, $Y_m$ is 
the fiber of $\caly\to M$ over $m\in M$, and we are associating codimension 
$k$ cycles with $(k,k)$-forms via Poincar\'e duality.) 

Before defining the normal function, we note that in the geometric 
situation, we have the isomorphism ($n:=\dim(Y)$)
\begin{equation}
\eqlabel{note}
J^{2k-1}(Y_m) = \bigl(F^{n-k+1} H^{2n-2k+1}(Y_m)\bigr)^*/H_{2n-2k-1}(Y_m;\zet) ,
\end{equation}
which follows from the isomorphism $H^{2k-1}/F^k H^{2k-1} \cong 
\bigl(F^{n-k+1}H^{2n-2k+1}\bigr)^*$, and the equivalence from Poincar\'e
duality, $\bigl( H^{2k-1}(Y;\zet)\bigr)^* \cong H_{2n-2k+1}(Y;\zet)$,
given by integration. 

Stepping on \eqref{note}, to define the normal function associated with
$\calc$, we need to specify a map $\nu_\calc: F^{n-k+1}
\calh^{2n-2k+1}\to \calo_M$, defined modulo periods $H_{2n-2k+1}(Y;\zet)$. 
To this end, for each $m\in M$, we pick a $2n-2k+1$-chain $\Gamma_m$, such 
that
\begin{equation}
\eqlabel{chainchoice}
\del\Gamma_m = C_m \qquad \text{in\;\; $Y_m$} ,
\end{equation}
where $C_m=\calc\cap Y_m$ as a codimension-$k$ cycle. Such a chain
exists because $C_m$ is homologically trivial, but is ambiguous by closed
$2n-2k+1$ cycles. If we require that $\Gamma_m$ depend in a continuous
fashion on $m$, the ambiguity is reduced to $H_{2n-2k+1}(Y;\zet)$.

Now given $[\omega]\in F^{n-k+1}\calh^{2n-2k+1}$, we can locally on $M$ 
represent it by a relative ${2n-2k+1}$-form $\omega\in F^{n-k+1} \cala^{2n-2k+1}$ 
that is closed in the fiber direction and well-defined up to the image
of $d^Y: F^{n-k+1} \cala^{2n-2k}\to F^{n-k+1} \cala^{2n-2k+1}$ (this last 
assertion follows from the Dolbeault theorem). We then define
\begin{equation}
\eqlabel{wellgiven}
\nu_\calc([\omega])_m := \int_{\Gamma_m} \omega_m .
\end{equation}

Let us check that this is well-defined. If we choose a different representative
$\omega'$ of $[\omega]$, the difference is
\begin{equation}
\eqlabel{final}
\int_{\Gamma_m}(\omega_m' - \omega_m) = \int_{\del\Gamma_m} \alpha_m ,
\end{equation}
where $\alpha\in F^{n-k+1}\cala^{2n-2k}$. This vanishes by type considerations
since $\del\Gamma_m=C_m$ is holomorphic, so Poincar\'e dual to a $(k,k)$-form.

Finally, we check holomorphicity and transversality. Namely, we analyze the
variation of \eqref{wellgiven} as $m$ varies to first order in $M$. 
If $v$ is a (not necessarily holomorphic) complexified tangent vector to 
$M$ at $m$, Kodaira--Spencer theory provides us with a lift, $v'$, of $v$ to 
$T\caly$. The differential of \eqref{wellgiven} in the direction of $v$ 
can be written as
\begin{equation} 
\eqlabel{toll}
(d_v\nu_\calc([\omega]))_m = - \int_{C_m} (\omega_m,v') + 
\int_{\Gamma_m}(\tilde\nabla_v\omega)_m ,
\end{equation}
where $\tilde\nabla_v\omega$ represents $\nabla_v[\omega]$, and $\nabla$ is 
the Gauss--Manin connection on $\calh^{2n-2k+1}$.

To check holomorphicity, we let $v$ be anti-holomorphic and $[\omega]$ be a
holomorphic section of $F^{n-k+1}\calh^{2n-2k+1}$. We then have that 
$\nabla_v[\omega]=0$ in $\calh^{2n-2k+1}$. In fact, $(\tilde\nabla_v\omega)_m= 
d^Y (\omega_m,v')$ by Kodaira--Spencer. Thus, \eqref{toll} vanishes, and 
$\nu_\calc$ is a holomorphic section of $\calj^{2k-1}$.

To show transversality, we take $v$ to be holomorphic. Note that the statement 
$\nabla_v\tilde\nu_\calc\in F^{k-1}\calh^{2k-1}$ is under the isomorphism 
$H^{2k-1}/F^{k-1}H^{2k-1}\cong \bigl(F^{n-k+2} H^{2n-2k+1}\bigr)^*$
(see \eqref{note}) equivalent to the assertion that $(\nabla_v\tilde\nu_\calc)([\omega])_m 
= 0$, for $[\omega]\in F^{n-k+2} \calh^{2n-2k+1}$, and where $\tilde\nu_\calc$ is any 
lift of $\nu_\calc$ to $\bigl(\calh^{2n-2k+1}\bigr)^*$. By the compatibility
of the Gauss--Manin connection with Poincar\'e duality,
$d_v(\tilde\nu_\calc([\omega]))_m = (\nabla_v\tilde\nu_\calc)([\omega])_m +
\tilde\nu_\calc(\nabla_v[\omega])_m$, so this criterion becomes 
$d\tilde\nu_\calc([\omega])_m = \tilde\nu_\calc(\nabla_v[\omega])_m $, which 
is already independent of the lift. Now if $[\omega]\in F^{n-k+2}\calh^{2n-2k+1}$, 
$(\omega_m,v')\in F^{n-k+1}\calh^{2n-2k}$, so the first term in \eqref{toll} vanishes 
by type consideration. This implies transversality.

We close this subsection with one more definition: The association 
\begin{equation}
\eqlabel{abeljacobi}
\AJ:\calz^k(\caly)_{\rm hom}\to \calj^{2k-1}(\caly)\,,\qquad \calc\mapsto\nu_\calc
\end{equation}
is known as the {\it Abel--Jacobi} map. It contains some useful 
information about algebraic cycles and their algebraic equivalences. 
The theory is particularly rich for Calabi--Yau threefolds (as mentioned
above, the interesting value is then $k=2$), and led to a lot
of early results on questions related to holomorphic curves 
\cite{griffiths1,clemensinf,voisininf}. That subject was later 
revolutionized by mirror symmetry and Gromov--Witten theory. As 
we will try to convey in this article, normal functions are returning
to the enterprise as well, with promising applications in the context 
of D-branes and mirror symmetry for open strings.

\subsection{Abel--Jacobi map on the derived category}

To explain the relevance of normal functions to D-branes in general,
we take as starting point Witten's holomorphic Chern--Simons
functional. We denote by $Y$ a (compact) Calabi--Yau threefold, 
$E$ a holomorphic vector bundle over $Y$, with $\delbar$ the
Dolbeault operator coupled to $E$. If $a\in A^{(0,1)}(Y,\End(E))$
is a $(0,1)$-form with values in the endomorphisms of $E$, we define
\begin{equation}
\eqlabel{hCS}
S_{\rm hCS}(a)= \int_Y 
\tr \Bigl(\frac 12 a\wedge\delbar a + \frac 13 a\wedge a\wedge a\Bigr)
\wedge\Omega ,
\end{equation}
where $\Omega$ is the (unique up to scale) holomorphic $(3,0)$-form
on $Y$. The functional \eqref{hCS} was originally proposed in 
\cite{wittenissues}, as an expression for the spacetime superpotential 
in the context of the heterotic string. This proposal can be explained 
on the basis that the critical points of \eqref{hCS} are precisely those 
$a\in A^{(0,1)}(Y,\End(E))$ for which the $(0,2)$-part of the curvature 
vanishes,
\begin{equation}
F^{(0,2)} = \delbar a + a\wedge a = 0 ,
\end{equation}
\ie, the deformed operator $\delbar_a=\delbar+a$ is an alternative
Dolbeault operator on $E$, viewed as a differentiable vector bundle
on $Y$. In general $\delbar_a$ will define a different complex
structure on $E$.

In \cite{wcs}, Witten showed that, in the context of the topological
string, the functional \eqref{hCS} is the target space or string
field theory action describing the tree-level dynamics of open
strings on $Y$ coupled to $E$ (a topological B-brane). This also 
led to the suggestion that holomorphic Chern--Simons should make 
sense as a quantum theory, and to various puzzles related to 
non-renormalizability of \eqref{hCS}, appearance of closed strings 
as intermediate states, \etc. The classical theory has been analyzed 
in depth over the years, see \cite{calinreview} for a review. The 
recent results on open-closed topological string \cite{extended} 
can be viewed as giving partial answers to the problems related 
to the quantum theory.

Connections of the holomorphic Chern--Simons functional with the theory 
of normal functions have appeared in the mathematical literature in
\cite{doth,clemens2}, see also \cite{tyurin}. In the physics 
literature, a relation to the Abel--Jacobi map for curves on 
Calabi--Yau threefold was established, \eg, in 
\cite{vafaextend,kklm2,diva1,agva}. When our B-brane, instead 
of being specified by a holomorphic vector bundle, is wrapping a 
holomorphic curve $C$, it was shown in \cite{kklm2,diva1,agva} that 
the dimensional reduction of the holomorphic Chern--Simons action
is nothing but the Abel--Jacobi integral
\begin{equation}
\eqlabel{mina}
S(C) = \int_\Gamma \Omega \quad\text{with\; $\del\Gamma= C-C_0$}
\end{equation}
viewed as a functional on all possible curves homotopic to some given
reference holomorphic curve $C_0$.

Neglecting the dynamics of open strings, the most direct physical 
interpretation of the formulas \eqref{hCS} and \eqref{mina}, is as 
the {\it tension of BPS domainwalls} connecting the background 
vacuum ($\delbar$ or $C_0$), on the D-brane worldvolume, with some 
other vacuum, corresponding to a non-trivial critical point, $a_*$ 
or $C_*$, respectively.
\begin{equation}
\eqlabel{domainwall}
\calt = \begin{cases} \calw(a_*)-\calw(0) = S_{\rm hCS}(a_*) \\
\calw(C_*)-\calw(C_0) = S(C_*) 
\end{cases}
\end{equation}
It should be clear that, even neglecting open string dynamics, those 
expressions cannot be fully satisfactory for describing the superpotential
for an arbitrary B-brane, which might be neither a holomorphic
vector bundle nor a holomorphic curve in general. The algebraic 
device needed to generalize these formulas to an arbitrary object 
$B$ in $D^b(Y)$ (or some category equivalent to it) is the notion 
of the {\it algebraic second Chern class}, $c_2^{\rm alg}(B)$
\cite{MR0116023}. 
It takes values in the Chow group ${\CH}^2(Y)$ of algebraic cycles 
of codimension $2$, modulo rational equivalence. The image of 
$c_2^{\rm alg}(B)$ in cohomology $H^4(Y;\zet)$ is equal to the ordinary 
(topological) second Chern class $c_2^{\rm top}(B)$, but $c_2^{\rm alg}$ 
is generally a more refined invariant. 

The algebraic Chern class satisfies axioms very similar to its topological
counterpart. In particular, it splits exact triangles in the D-brane 
category. If
\begin{equation}
\xymatrix@R=1cm@C=0.7cm{
& B \ar[dr]^{} & \\
\ar[ur]^{} A && \ar[ll]_{[1]} C 
}
\end{equation}
for three objects, $A$, $B$, and $C$, then
\begin{equation}
c^{\rm alg}(A) - c^{\rm alg}(B) + c^{\rm alg}(C) = 0
\end{equation}
which together with functoriality (and its behavior on holomorphic
line bundles) is essentially enough to define
$c^{\rm alg}$.

Using the algebraic second Chern class puts us directly in the situation 
discussed in the previous subsection. When $c_2^{\rm top}(B)=0\in H^4(Y;\zet)$, 
the algebraic cycle defined by $c_2^{\rm alg}(B)\in \CH^2(Y)$ is 
homologically trivial, and yields a normal function
$\nu_B= \nu_{c_2^{\rm alg}(B)}$. In particular the formula
for the domainwall tension is
\begin{equation}
\calt = \nu_{B} (\Omega) ,
\end{equation}
where $\Omega$ is the same holomorphic three-form as above.
It is not hard to see that this definition reduces to \eqref{hCS}
and \eqref{mina} when $B$ is a holomorphic vector bundle or a
holomorphic curve, respectively. 

We should emphasize that the second Chern class will certainly not 
capture all the intricacies of the superpotential for a general B-brane 
on a Calabi--Yau. This will require a much more sophisticated analysis,
partly along the lines of the cited literature.

\subsection{Comments on open problems}

Before turning to the applications, we will collect a few more
remarks from the general theory of normal functions, some
of which might prove valuable for further developments.

\paragraph{Extension of Hodge structure}

Let $(H_\zet^{2k-1},F^*\calh^{2k-1})$ be an integral variation of Hodge
structure of weight $2k-1$ over a base $M$. Let $\nu$ be a normal
function. In exercises 1 and 2 in Chapter 7 of \cite{voisinbook2}, 
it is shown that this data can be used to define an extension of Hodge
structure to yield an integral variation of {\it mixed} Hodge
structure. At the integral level, this is a locally trivial extension
\begin{equation}
\eqlabel{extension}
H_\zet^{2k-1} \to (H_\zet= H_\zet^{2k-1}\oplus\zet) \to \zet .
\end{equation}
The weight filtration is given by $W_{2k-2}H_\zet=0$, 
$W_{2k-1}H_\zet=H_\zet^{2k-1}$, $W_{2k}=H_\zet$, while
the Hodge filtration on $\calh=(H_\zet^{2k-1}\oplus\zet)\otimes
\calo_M$ is such that it reduces to the given Hodge filtration  
$F^*\calh^{2k-1}$ on $\calh^{2k-1}$, and to $F^{k+1}\calo_M=0$, 
$F^k\calo_M=\calo_M$ on the quotient.

In the context of mirror symmetry, a {\it different} mixed Hodge structure
is relevant. This mixed Hodge structure is associated with the degeneration 
at a point of maximal unipotent monodromy in the moduli space 
\cite{guide,delignelimit}. The monodromy calculations of \cite{open}, 
partially reviewed in section \ref{main} are indicative of a very interesting 
interaction between this limiting mixed Hodge structure and the one given 
by extension using the normal function \eqref{extension}. It would be 
interesting to elucidate this further.

\paragraph{More extensions?}

We have so far largely suppressed the existence of an $A_\infty$-structure
on the category of B-branes, except to ask the natural question how much
of that structure is possibly captured by the normal function. 
In thinking about this problem, we are led to the following speculations.

The $A_\infty$-structure on a brane $B$ in the category of B-branes is
given by a collection of ``higher'' products $m_n$ satisfying certain
conditions of associativity. At the level of the string worldsheet,
the $m_n$ with $n\ge 2$ can be determined by computing the (topological) 
disk amplitudes with $n+1$ open string insertions on the boundary. $m_1$
is identified with the open string BRST operator. Finally, $m_0$ is related
to the bulk-to-boundary obstruction map by taking one derivative with
respect to the closed string moduli \cite{extended}.

From general considerations, as well as the identification of the disk 
amplitude with two bulk insertions as the Griffiths infinitesimal invariant 
\cite{extended}, it appears natural that the normal function $\nu$ can 
fit as an ``$m_{-1}$'' into the $A_\infty$-structure.
As emphasized in \cite{extended}, the obstruction map can be interpreted
Hodge theoretically as the dual of the infinitesimal Abel--Jacobi map.
Those two observations suggest that one should try to understand
whether the higher $A_\infty$ products $m_n$ for $n\ge 1$ can also 
be given a Hodge theoretic interpretation.

\paragraph{A-model version}

All considerations in this paper are phrased in the language of the
B-model. On the other hand, we recall that much of the deeper
understanding of classical (closed string) mirror symmetry involved
the reconstruction of Hodge theoretic structures in the A-model.
In particular, the importance of quantum cohomology and the 
structure of the mirror map become especially clear in the ``A-model 
variation of Hodge structure'' \cite{compact,parkcity}.

It would be very interesting to extend these insights to
the open string. A general definition of a functional conjecturally 
mirror to the holomorphic Chern--Simons functional/domainwall tension 
(see, eq.\ \eqref{domainwall}) is given in \cite{psw}, extending 
\cite{wcs}. This functional includes corrections from worldsheet 
instantons (holomorphic disks ending on Lagrangian submanifolds), 
and should in principle be related to Floer theory and the Fukaya 
category, as the open string analogues of quantum cohomology. 
This relation should be similar to that between the holomorphic Chern--Simons 
functional and the derived category. In the A-model, the precise relation 
is not currently understood, but as an intermediate step, it would 
be interesting to check at least the Hodge theoretic statements 
pertaining to normal functions, based on, say, axioms for open 
Gromov--Witten invariants.

\section{The Real Quintic and its Mirror}
\label{realquintic}

Our interest now turns to the quintic Calabi--Yau $X=\{G=0\}\subset\projective^4$,
defined as the vanishing locus of a degree $5$ polynomial $G$ in 5
complex variables $x_1,\ldots,x_5$. We assume that $X$ is defined over
the reals, which means that all coefficients of $G$ are real (possibly up 
to some common phase). The real locus $\{x_i=\bar x_i\}\subset X$
is then a Lagrangian submanifold, and after choosing a flat $U(1)$ connection, 
will define an object in the (derived) Fukaya category $\fuk(X)$. 
In this section, we will first review a proposal which identifies a mirror 
object in the category of B-branes of the mirror quintic, in its 
Landau--Ginzburg description. Via some detours, we will be able to derive 
from the matrix factorization the corresponding normal function. In the next 
section, we will then show by an explicit computation that this normal 
function satisfies precisely the inhomogeneous Picard--Fuchs equation proposed 
in \cite{open}.

\subsection{625 real quintics}

Both the topological type and the homology class in $H_3(X;\zet)$ of the real 
locus depend on the complex structure of $X$ (the choice of (real) polynomial 
$G$). On the other hand, the Fukaya category is independent of the choice
of $G$ (real or not). The object in $\fuk(X)$ that we shall refer to 
as the real quintic is defined from the real locus $L$ of $X$ when $G$ is
the Fermat quintic $G=x_1^5+x_2^5+x_3^5+x_4^5+x_5^5$. It is not
hard to see that topologically, $L\cong \reals\projective^3$. There
are therefore two choices of flat bundles on $L$, and we will denote
the corresponding objects of $\fuk(X)$ by $L_+$ and $L_-$, respectively.
More precisely, since $\fuk(X)$ depends on the choice of a complexified
K\"ahler structure on $X$, we define $L_\pm$ for some choice of K\"ahler
parameter $t$ close to large volume $\Im(t)\to\infty$, and then continue
it under K\"ahler deformations. In fact, the rigorous definition of the Fukaya 
category is at present only known infinitesimally close to this large 
volume point \cite{fooo}. However, $\fuk(X)$ 
does exist over the entire stringy K\"ahler moduli space of $X$, and
at least some of the structure varies holomorphically. Our interest here 
is in the variation of the categorical structure associated with $L_\pm$ 
over the entire stringy K\"ahler moduli space of $X$, identified via
mirror symmetry with the complex structure moduli space of the mirror
quintic, $Y$.

The Fermat quintic is invariant under more than one anti-holomorphic
involution. If $\zet_5$ denotes the multiplicative group of fifth roots
of unity, we define for $\chi=(\chi_1,\ldots,\chi_5)\in (\zet_5)^5$ an
anti-holomorphic involution $\sigma_\chi$ of $\projective^4$ by its action 
on homogeneous coordinates
\begin{equation}
\eqlabel{actionon}
\sigma_\chi : x_i \to \chi_i \bar x_i .
\end{equation}
The Fermat quintic is invariant under any $\sigma_\chi$. The involution and
the fixed point locus only depend on the class of $\chi$ in 
$(\zet_5)^5/\zet_5\cong (\zet_5)^4$, and we obtain in this way 
$5^4=625$ (pairs of) objects $L^{[\chi]}_\pm$ in $\fuk(X)$. We will
return to those $625$ real quintics below, and for the moment focus
on $L_\pm = L^{[\chi=1]}_\pm$.

We emphasize again that although we have defined the Lagrangians 
$L_\pm^{[\chi]}$ as fixed point sets of anti-holomorphic involutions
of the Fermat quintic, we can think of the corresponding objects of 
${\rm Fuk}(X)$ without reference to the complex structure.

\subsection{The prediction}

The image in $K^0(\fuk(X))$ is the same for $L_+$ and $L_-$. This is
the counterpart in the A-model of the triviality of topological Chern 
classes $c^{\rm top}(B_+-B_-)$ for two objects $B_\pm$ in the category of
B-branes. As mentioned above, there should exist a definition of an 
Abel-Jacobi map to a normal function of the A-model variation of 
Hodge structure constructed from the quantum cohomology of $X$ 
\cite{compact,parkcity}. As explained in \cite{open},
this normal function can be realized geometrically by wrapping a D-brane 
on a disk $D$ whose boundary on $L$ represents the non-trivial element of 
$H_1(L;\zet)\cong\zet_2$. Neglecting instanton corrections, the corresponding 
truncated normal function is $\frac t2 \pm \frac 14 \bmod t \zet + \zet$.%
\footnote{The sign depends on whether we consider $L_+-L_-$ or 
$L_--L_+$. For this to make sense, note that $\frac t2 - \frac 14=
-(\frac t2 +\frac 14) \bmod t\zet+\zet$. For details, see \cite{open}.} 
Instanton corrections deform this to
\begin{equation}
\eqlabel{deform}
\calt_A(t) = \frac{t}{2} \pm \Bigl( \frac 14 + \frac{1}{2 \pi^2} 
\sum_{d\;{\rm odd}} n_d q^{d/2} \Bigr) ,
\end{equation}
where $q=\ee^{2\pi\ii t}$, and $n_d$ are the open Gromov--Witten invariants
of the real quintic defined in \cite{solomon}, predicted in \cite{open},
and fully computed in \cite{psw}. The precise result for the $n_d$ is as 
follows.

Mirror symmetry for the quintic is governed by the differential operator
\begin{equation}
\eqlabel{pfop}
\call= \theta^4 - 5z (5\theta+1)(5\theta+2)(5\theta+3)(5\theta+4)\,,
\end{equation}
where $\theta = z d/dz$. As we will review further below, $\call$ is the 
Picard--Fuchs operator of the mirror quintic. The equation $\call\varpi(z)=0$
has four linearly independent solutions. Two of those solutions are
given by the following power-series expansion around $z=0$:
\begin{equation}
\eqlabel{fundamental}
\begin{split}
\varpi_0(z) &= \sum_{m=0}^\infty \frac{(5m)!}{(m!)^5} z^m \\
\varpi_1(z) &= \varpi_0(z) \log z + 5 \sum_{m=1}^\infty
\frac{(5m)!}{(m!)^5} z^m\bigl[\Psi(1+5m)-\Psi(1+m)\bigr] \\
\end{split}
\end{equation}
and determine the mirror map as
\begin{equation}
t = t(z) = \frac{1}{2\pi\ii} \frac{\varpi_1(z)}{\varpi_0(z)}\,,\qquad 
q(z) = \exp(2\pi\ii t(z)) .
\end{equation}
The result of \cite{open,psw} is
\begin{equation}
\eqlabel{inhomopf}
\call \bigl(\varpi_0(z)\calt_A(z)\bigr) = \frac{15}{16\pi^2} \sqrt{z} .
\end{equation}
Combined with the boundary conditions \eqref{deform}, this is equivalent to
\begin{equation}
\eqlabel{lvasym}
\varpi_0(z)\calt_A(z) = \frac{\varpi_1(z)}{4\pi\ii}+ \frac{\varpi_0(z)}{4} + 
\frac{15}{\pi^2}\tau(z)
\end{equation}
where
\begin{equation}
\tau(z) = \frac{\Gamma(3/2)^5}{\Gamma(7/2)}\;
\sum_{m=0}^\infty \frac{\Gamma(5m+7/2)}{\Gamma(m+3/2)^5}\; z^{m+1/2}
= \sqrt{z} + \frac{5005}{9} z^{3/2} + \cdots 
\end{equation}
gives a particular solution of the inhomogeneous Picard--Fuchs equation
\eqref{inhomopf}.

\subsection{Matrix factorization}

For the rest of this work, $W$ will denote the one-parameter family of 
quintic polynomials
\begin{equation}
\eqlabel{onepar}
W = \frac 15\bigl(x_1^5+x_2^5+x_3^5+x_4^5+x_5^5\bigr) -\psi x_1x_2x_3x_4x_5 .
\end{equation}
Geometrically, the mirror quintic, $Y$, is the quotient of this one-parameter 
family of quintics by $(\zet_5)^3=(\zet_5)^4/\zet_5$, where $(\zet_5)^4$ is
the group of phase symmetries of $W$ (for $\psi\neq 0$). Alternatively, we 
can think of a Landau--Ginzburg orbifold model with worldsheet superpotential
$W$ and orbifold group $(\zet_5)^4$. 

We will also have occasion to work in the B-model on the one-parameter family 
of quintic hypersurfaces given by $W=0$ in $\projective^4$ (without quotient). 
In this context, we will denote this family by $X_\psi$. When we work in the
context of the A-model, with an arbitrary complex structure represented
by a general quintic polynomial $G$, we will 
continue to denote the quintic simply by $X$.

Recall that for a quintic $X$ defined by $G=0$, the homological Calabi--Yau/Landau--Ginzburg 
correspondence \cite{stability,orlov,Aspinwall:2006ib,hhp} states that the 
derived category of coherent sheaves of $X$ is equivalent to the graded, 
equivariant category of matrix factorizations of the corresponding 
Landau--Ginzburg superpotential,
\begin{equation}
\eqlabel{cylgquintic}
D^b(X) = \MF(G/\zet_5) ,
\end{equation}
where $\zet_5$ is the diagonal group of phase symmetries. The analogous
statement for 
the mirror quintic is
\begin{equation}
\eqlabel{cylgmirror}
D^b(Y) \cong \MF(W/\zet_5^4) ,
\end{equation}
with $(\zet_5)^4$ as above.

To describe an object mirror to the real quintic, we begin with finding a
matrix factorization of the one-parameter family of superpotentials \eqref{onepar}.
If $V\cong \complex^5$ is a $5$-dimensional vector space, we can associate to 
its exterior algebra a $\complex[x_1,\ldots,x_5]$-module $M=\wedge^* V\otimes
\complex[x_1,\ldots,x_5]$. It naturally comes with the decomposition
\begin{equation}
\eqlabel{decomp}
M = M_0+M_1+M_2+M_3+M_4+M_5\,, \qquad \text{where $M_s=\wedge^sV\otimes 
\complex[x_1,\ldots,x_5]$} ,
\end{equation}
and the $\zet_2$-grading $(-1)^i$. Let $\eta_i$ be a basis of $V$ and 
$\bar\eta_i$ the dual basis of $V^*$, both embedded in ${\rm End}(M)$. We 
then define two families of matrix factorizations $(M,Q_{\pm})$ of $W$ by
\begin{equation}
\eqlabel{deform1}
Q_{\pm} = \frac{1}{\sqrt{5}}
\sum_{i=1}^5 (x_i^2\eta_i+x_i^3\bar\eta_i) \pm \sqrt{\psi} 
\prod_{i=1}^5 (\eta_i -x_i\bar \eta_i) .
\end{equation}
To check that $Q_\pm^2=W\cdot {\rm id}_M$, one uses that $\eta_i$,
$\bar\eta_i$ satisfy the Clifford algebra
\begin{equation}
\eqlabel{clifford}
\{\eta_i,\bar\eta_j\} = \delta_{ij} ,
\end{equation}
as well as the ensuing relations
\begin{equation}
\{(x_i^2\eta_i + x_i^3\bar\eta_i),(\eta_i-x_i\bar\eta_i)\} = 0 
\qquad
{\rm and}
\qquad
(\eta_i-x_i\bar\eta_i)^2 = -x_i .
\end{equation}
The matrix factorization \eqref{deform1} is quasi-homogeneous ($\complex^*$-gradable). 
The R-charges of the superpotential and the $x_i$ are $2$ and $2/5$, respectively. So
if we assign R-charge $1/5$ and $-1/5$ to $\eta_i$ and $\bar\eta_i$, respectively,
$Q$ will have uniform R-charge $1$. Since $Q$ is irreducible, this determines the 
R-charge of $M$ uniquely up to an overall shift. As explained in \cite{stability},
this ambiguity should be fixed by $\tr R=0$ for studying the stability of 
the matrix factorizations. But for the present purposes, we will use a different
convention, see below.

To specify objects in $\MF(W/\Gamma)$, where $\Gamma=\zet_5$ or $(\zet_5)^4$
for the quintic and mirror quintic, respectively, we have to equip $M$ with
a representation of $\Gamma$ such that $Q$ is equivariant with respect to the 
action of $\Gamma$ on the $x_i$. Since $Q$ is irreducible, this representation
of $\Gamma$ on $M$ is determined up to a character of $\Gamma$ by a representation 
on $V$, \ie, an action on the $\eta_i$. For $\gamma\in\Gamma$, we have $\gamma(x_i) 
= \gamma_i x_i$ for some fifth root of unity $\gamma_i$. We then set $\gamma(\eta_i)=
\gamma_i^{-2} \eta_i$, making $Q$ equivariant. As noted, this representation is
unique up to an action on $M_0$, \ie, a character of $\Gamma$.

For the mirror quintic, $\Gamma = {\rm Ker}((\zet_5)^5\to\zet_5)$, so 
$\Gamma^* = (\zet_5)^5/\zet_5$, and we label its characters as $[\chi]$. 
The corresponding objects of $\MF(W/\Gamma)$ constructed out of $Q_\pm$ \eqref{deform1} 
are classified as $Q_\pm^{[\chi]}=(M,Q_\pm,\rho_{[\chi]})$, where $\rho_{[\chi]}$ is the 
representation on $M$ we just described.

\medskip

{\bf Conjecture:} There is an equivalence of categories $\fuk(X)\cong \MF(W/(\zet_5)^4)$
which identifies the 625 pairs of objects $L_\pm^{[\chi]}$ with the 625 pairs of 
equivariant matrix factorizations $Q_\pm^{[\chi]}$.

\medskip

{\it Note:} One can formulate a similar conjecture for any hypersurface in weighted 
projective space which has a Fermat point in its complex structure moduli space.

\subsection{Intersection Index}

The first piece of evidence for the above conjecture comes from ref.\ \cite{bdlr}.
In that paper, the $625$ Lagrangian submanifolds of $X$ described above were associated
with the so-called ${\bf L}=(1,1,1,1,1)$ A-type Recknagel--Schomerus states in the Gepner 
model. These A-type boundary states had been constructed in \cite{resc} as tensor products 
of Cardy states in the $\caln=2$ minimal model building blocks of the Gepner model. In 
turn, these Cardy states of the minimal model were identified in \cite{hiv} with the 
Lagrangian wedge branes of opening angle $4\pi/5$ in the Landau--Ginzburg description of the 
$\caln=2$ minimal models. Via mirror symmetry for the minimal models, those wedges are 
equivalent to the matrix factorizations based on $x_i^5=x_i^2\cdot x_i^3$, see, \eg, 
\cite{horiminimal}. These are precisely the building blocks of the factorization 
\eqref{deform1}, specialized to $\psi=0$. The above deformation away from $\psi=0$,
as well as the identification of the pairs $Q_\pm^{[\chi]}$ with the pairs of objects
$L_\pm^{[\chi]}$ was first noted in \cite{howa}, following the suggestion of \cite{bdlr}.

The initial step in the above identification of $L_\pm^{[\chi]}$ with $Q_\pm^{[\chi]}$ 
was justified in \cite{bdlr} by a comparison of the intersection indices of the 
$L_\pm^{[\chi]}$ with the corresponding intersection indices of the Gepner model boundary
states. We will reproduce this here using the matrix factorizations. The match of the 
domainwall tensions\footnote{Note that because of the symmetries,
these domainwall tensions do not depend on the discrete group representation.} $L_+-L_-$ and $Q_+-Q_-$ computed in 
the A- and B-model, respectively, constitutes further evidence for the above 
conjecture. 

Let us start with the geometric intersection index between\footnote{The 
intersection index, being topological, does not depend on the 
Wilson lines on the A-branes. For the B-branes, it is correspondingly independent of the 
sign of the square root in \eqref{deform1}.}
$L^{[\chi]}$ and
$L^{[\chi']}$. Because of the projective equivalence,
we have to look at the intersection of the fixed point loci of $\sigma_\chi$
and $\sigma_{\omega\chi'}$ from \eqref{actionon} where $\omega$ runs over the $5$ fifth 
roots of unity. It is not hard to see that topologically
\begin{equation}
{\rm Fix}(\sigma_{\chi})\cap {\rm Fix}(\sigma_{\omega\chi'})\cap X \cong 
\reals\projective^{d-2}\,, \qquad\text{where $d=\# \{\chi'_i=\om\chi_i\}$} .
\end{equation}
After making the intersection transverse by a small deformation in the normal direction,
we obtain a vanishing contribution for $d=0,1,3,5$, and $\pm 1$ for $d=2,4$, where the
sign depends on the non-trivial phase differences $\chi_i^* \om\chi'_i$. Explicitly,
one finds
\begin{equation}
\eqlabel{inter1}
L^{[\chi]}\cap L^{[\chi']} = \sum_{\om\in\zet_5} f_1(\chi'^*\om\chi) ,
\end{equation}
where
\begin{equation}
f_1(\chi) = 
\begin{cases} \prod_{i=1}^5 {\rm sgn}\bigl( \Im(\chi_i)\bigr)\,, & \text{if $\#\{i,\chi_i=1\}=2,4$} \\
0 & \text{else} .
\end{cases}
\end{equation}
To compute the intersection index between the matrix factorizations, we use the index
theorem of \cite{stability}. It says in general
\begin{multline}
\eqlabel{indextheorem}
\chi\Hom\bigl((M,Q,\rho),(M',Q',\rho')\bigr) :=
\sum_{i} (-1)^i \dim\Hom^i\bigl((M,Q,\rho),(M',Q',\rho')\bigr)=\\
\frac{1}{|\Gamma|} \sum_{\gamma\in\Gamma} \str_{M'} \rho'(\gamma)^*
\frac{1}{\prod_{i=1}^5(1-\gamma_i)}\str_M \rho(\gamma) ,
\end{multline}
where $\gamma_i$ are the eigenvalues of $\gamma\in\Gamma$ acting on the $x_i$,
and $\rho$, $\rho'$ are the representations of $\Gamma$ on $M$. For $M=M'$, $Q=Q'$ and 
$\rho=\rho_{[\chi]}$, $\rho'=\rho_{[\chi']}$ described above, this evaluates to
\begin{equation}
\eqlabel{inter2}
-\frac{1}{5^4} \sum_{\gamma\in(\zet_5)^4} \chi(\gamma')^*\chi(\gamma)
\prod_{i=1}^5 (\gamma_i+\gamma_i^2-\gamma_i^3-\gamma_i^4)
= - \sum_{\om\in\zet_5} f_2(\chi'^*\om\chi) ,
\end{equation}
where
\begin{equation}
f_2(\chi) = \begin{cases}
\prod_{i=1}^5 {\rm sgn}\bigr(\Im(\chi_i)\bigr)\,, & \text{if $\#\{i,\chi_i=1\} = 0$} \\
0 & \text{else} .
\end{cases}
\end{equation}
We do not know any generally valid result from the representation theory
of finite cyclic group which shows that \eqref{inter1} and \eqref{inter2}
coincide. It is however not hard to check by hand or computer that for
all $\chi$,
\begin{equation}
\sum_{\om\in\zet_5} (f_1+f_2)(\om\chi) = 0 .
\end{equation}
Hence
\begin{equation}
L^{[\chi]}\cap L^{[\chi']} = \chi\Hom(Q^{[\chi]},Q^{[\chi']})
\end{equation}
as claimed.

\subsection{Bundles}

We now proceed with the construction of the normal function from the matrix
factorization \eqref{deform1}. To this end, we use the homological Calabi--Yau/Landau--Ginzburg
correspondence \eqref{cylgquintic} for the quintic as described in \cite{hhp}. 
This will produce for us a set of $5$ complexes of coherent 
sheaves (bundles) on the one-parameter family of quintics $X_\psi$. By making those 
equivariant with respect to the geometric $(\zet_5)^3$ action, this will yield the 
$625$ objects in $D^b(Y)$ mirror to the real quintics. 

The technique underlying the algorithm of \cite{hhp} is the gauged linear sigma
model of \cite{Witten:1993yc}. Thus, we first construct a D-brane in the gauged
linear sigma model from the equivariant matrix factorization, and in the second
step a complex of (line) bundles on the quintic. We have to and can live with
two ambiguities in the construction. The first  ambiguity is the Landau--Ginzburg 
monodromy (cyclic permutation of the characters of $\Gamma=\zet_5$), while the second 
depends on a certain ``Band Restriction Rule'' for assignment of the gauge charges in
the linear sigma-model. The upshot of the construction is the following. We can view
the matrix factorization, namely, the $\zet_2$-graded module $M$ equipped with
$Q$ of $Q^2=W$, as a $2$-periodic infinite complex over the affine singularity
$W=0$. We then {\it truncate} this infinite complex to a {\it semi-infinite complex}
in a way that depends on the charge and representation assignments in the gauged
linear sigma model. The departure of this construction from the traditional (Serre)
correspondence between sheaves on the hypersurface and graded modules on the affine
singularity is that the cohomological grading of the complexes also depends on the
linear sigma model charges. We now implement this algorithm in our example, 
referring to \cite{hhp} for the complete details.

Given $(M,Q,\rho_\chi)$, we first assign R-charges (\ie, a $\complex^*$-representation, 
generated by a rational Hermitian matrix, $R$ on $M$ in such a way that
\begin{equation}
\ee^{\ii\pi R} = \rho_\chi(\gamma) (-1)^s ,
\end{equation}
where $(-1)^s$ is the $\zet_2$-grading on $M$, and $\gamma\equiv \ee^{2\pi\ii/5}$ is 
the generator of $\zet_5$. In the decomposition \eqref{decomp}, $\rho_\chi(\gamma)=
\ee^{2\pi\ii(n-2s)/5}$, where $\chi=\ee^{2\pi\ii n/5}$. We choose the R-charge
assignment of $M_s$ in \eqref{decomp} as $R_s=\frac s5+\frac{2n}5$.

Following the algorithm of \cite{hhp} we now select a ''band'' of $5$ consecutive 
integers $\Lambda=\{0,1,2,3,4\}$ and find for each $s$ an integer $\tilde R_s=s\bmod 2$ 
and an integer $q_s\in\Lambda$ such that
\begin{equation}
R_s = \tilde R_s - \frac{2q_s}5 .
\end{equation}
$\tilde R_s$ and $q_s$ are uniquely determined by this equation. Depending on $n$, we
find for the pairs $(\tilde R_s,q_s)$ the following table
\begin{equation}
\begin{array}{|c|c|c|c|c|c|}
\hline
\text{\backslashbox{$s$}{$n$}} & 0   & 1     & 2     & 3     & 4     \\\hline
0 & (0,0) & (2,4) & (2,3) & (2,2) & (2,1) \\
1 & (1,2) & (1,1) & (1,0) & (3,4) & (3,3) \\
2 & (2,4) & (2,3) & (2,2) & (2,1) & (2,0) \\
3 & (1,1) & (1,0) & (3,4) & (3,3) & (3,2) \\
4 & (2,3) & (2,2) & (2,1) & (2,0) & (4,4) \\
5 & (1,0) & (3,4) & (3,3) & (3,2) & (3,1) \\\hline
\end{array}
\end{equation}
This data yields a graded, gauge invariant matrix factorization, $Q_{\rm GLSM}$ of the 
linear sigma model superpotential $W_{\rm GLSM}=P W$, where $P$ is Witten's P-field
\cite{Witten:1993yc}. In reducing to the non-linear sigma-model on the hypersurface,
the bulk modes of $P$ are integrated out, while the quantization of the single
boundary degree of freedom yields the Fock space of a harmonic oscillator, 
$\calh^P\cong \oplus_{N\ge 0} \calh^P_N$, where each $\calh^P_N\cong\complex$. 
The resulting complex on the quintic hypersurface is built from the tensor
product $M\otimes \calh^P$, where $M_s\otimes\calh^P_N$ is placed in homological
degree $d=\tilde R_s+2N$ and twisted by the line bundle $\calo(q_s+5N)$.
The original matrix factorization $Q$ acts on this complex in a way compatible
with all gradings.

For the data above, we obtain explicitly the following five complexes, 
(here, $V^s\equiv\wedge^s V$, and the integer in square brackets indicates the
homological degree of the first term in the complex).
\begin{equation}
\begin{split}
n=&0:[0]\\&
\begin{array}{c}
\calo(0)\otimes V^0\\
\phantom{\calo}
\\
\phantom{\calo}
\end{array}
\to
\begin{array}{c}
\calo(2)\otimes V^1\\
\calo(1)\otimes V^3 \\
\calo(0)\otimes V^5 
\end{array}
\to
\begin{array}{c}
\calo(5)\otimes V^0\\
\calo(4)\otimes V^2\\
\calo(3)\otimes V^4\\
\end{array}
\to
\begin{array}{c}
\calo(7)\otimes V^1\\
\calo(6)\otimes V^3 \\
\calo(5)\otimes V^5 
\end{array}
\to
\cdots\cdots 
\end{split}
\end{equation}
\begin{equation}
\begin{split}
n=&1:[1]\\&
\begin{array}{c}
\calo(1)\otimes V^1\\
\calo(0)\otimes V^3 \\
\phantom{\calo}
\end{array}
\to
\begin{array}{c}
\calo(4)\otimes V^0\\
\calo(3)\otimes V^2\\
\calo(2)\otimes V^4\\
\end{array}
\to
\begin{array}{c}
\calo(6)\otimes V^1\\
\calo(5)\otimes V^3 \\
\calo(4)\otimes V^5 
\end{array}
\to
\begin{array}{c}
\calo(9)\otimes V^0\\
\calo(8)\otimes V^2\\
\calo(7)\otimes V^4\\
\end{array}
\to \cdots\cdots\\
\end{split}
\end{equation}
\begin{equation}
\begin{split}
n=&2:[1]\\&
\begin{array}{c}
\calo(0)\otimes V^1\\
\phantom{\calo}
\\
\phantom{\calo}
\end{array}
\to
\begin{array}{c}
\calo(3)\otimes V^0\\
\calo(2)\otimes V^2\\
\calo(1)\otimes V^4\\
\end{array}
\to
\begin{array}{c}
\calo(5)\otimes V^1\\
\calo(4)\otimes V^3 \\
\calo(3)\otimes V^5 
\end{array}
\to
\begin{array}{c}
\calo(8)\otimes V^0\\
\calo(7)\otimes V^2\\
\calo(6)\otimes V^4\\
\end{array}
\to
\cdots\cdots 
\end{split}
\end{equation}
\begin{equation}
\eqlabel{neq3}
\begin{split}
n=&3:[2]\\&
\begin{array}{c}
\calo(2)\otimes V^0\\
\calo(1)\otimes V^2\\
\calo(0)\otimes V^4\\
\end{array}
\to
\begin{array}{c}
\calo(4)\otimes V^1\\
\calo(3)\otimes V^3 \\
\calo(2)\otimes V^5 
\end{array}
\to
\begin{array}{c}
\calo(7)\otimes V^0\\
\calo(6)\otimes V^2\\
\calo(5)\otimes V^4\\
\end{array}
\to
\begin{array}{c}
\calo(9)\otimes V^1\\
\calo(8)\otimes V^3 \\
\calo(7)\otimes V^5 
\end{array}
\to \cdots\cdots\\
\end{split}
\end{equation}
\begin{equation}
\begin{split}
n=&4:[2]\\&
\begin{array}{c}
\calo(1)\otimes V^0 \\
\calo(0)\otimes V^2\\
\phantom{\calo}
\end{array}
\to
\begin{array}{c}
\calo(3)\otimes V^1\\
\calo(2)\otimes V^3 \\
\calo(1)\otimes V^5 
\end{array}
\to
\begin{array}{c}
\calo(6)\otimes V^0\\
\calo(5)\otimes V^2\\
\calo(4)\otimes V^4\\
\end{array}
\to
\begin{array}{c}
\calo(8)\otimes V^1\\
\calo(7)\otimes V^3 \\
\calo(6)\otimes V^5 
\end{array}
\to
\cdots\cdots 
\end{split}
\end{equation}
The differential on these complexes is $Q$ from \eqref{deform1}, where
as before $\eta_i$ and $\bar\eta_i$ act on the exterior algebra $\wedge^* V$ in
the usual way. It would be interesting to obtain a more intrinsic description
of these 5 objects in $D^b(X_\psi)$, understand their deformations to a general
quintic,\footnote{We thank Tony Pantev and Ron Donagi for
extensive discussions on possible such descriptions.} investigate stability
at large volume, etc. It is not hard to compute the topological Chern characters 
of these five objects, and to check that they agree with those determined from 
\cite{bdlr}. For example, the virtual ranks of the objects are given by 
$(-3,3,-7,8,7)$ for $n=(0,1,2,3,4)$, respectively. The simplest and most 
canonical object appears to be the one corresponding to $n=3$. Namely, as 
found in \cite{bhhw}, it carries precisely the topological charges
required for anomaly cancellation in a type I (or type IIB orientifold) string
compactification on the quintic with non-trivial discrete B-field. (This is
mirror to a type IIA orientifold compactification on the mirror quintic.) It
is natural therefore to assume that this corresponds to a rank $8$ bundle
which moreover is stable at large volume on the quintic.

\subsection{From matrix factorization to curve}

The five complexes in the previous subsection define $5$ objects in $D^b(X_\psi)$.
(Although semi-infinite, they are quasi-isomorphic to finite complexes
because of the eventual periodicity.) As discussed before, to obtain
the $625$ objects in $D^b(Y)$ mirror to the real quintics, we have to
make these objects $(\zet_5)^3$ equivariant. It would be interesting to
understand this construction in detail, and in particular, what happens under
the resolution of the orbifold singularities. For our purposes however, 
we do not need this. In fact, to compute the normal function by
the Abel--Jacobi map, we do not even need to distinguish between the five
objects on the quintic. Note that the defining semi-infinite complexes 
differ only in low homological degree by extensions by line bundles, which
contribute only trivially to algebraic K-theory and the Abel--Jacobi map. 
In other words, {\it all the information about the normal function is 
contained in the $2$-periodic part of the complexes, which is nothing but 
the original matrix factorization!} This fact would have allowed us to 
bypass all the complications associated with the homological 
Calabi--Yau/Landau--Ginzburg correspondence. We nevertheless presented the 
detailed results in the previous subsection, because we feel that they might 
be of independent interest, for instance for questions of stability.

In this subsection, we proceed with the computation of the algebraic
second Chern classes of $Q_\eps^{[\chi]}$, where $\eps=\pm 1$. Specifically, the
domainwall tension of our interest is given by the image under the Abel--Jacobi 
map of $Q_+^{[\chi]}-Q_-^{[\chi]}$. Note that this is well-defined since, as 
follows \eg, from the index theorem \eqref{indextheorem}, the topological 
Chern classes only depend on $\chi$, and not on $\eps$, which is the sign
of the square root in \eqref{deform1}. On the other hand, the Abel--Jacobi
map is independent of $\chi$, as explained in the previous paragraph.

It does, however, make a difference whether we work on the quintic or its 
mirror. On the quintic, we can work with the explicit bundle representatives
from \eqref{neq3}. Let
\begin{equation}
E_\pm = {\rm Ker}\bigl(\calo(2)\oplus\calo(1)^{10}\oplus\calo(0)^5
\overset{Q_{\pm}}{\longrightarrow} \calo(4)^5\oplus\calo(3)^{10}\oplus\calo(2)\bigr) .
\end{equation}
In general, for a bundle of rank $r$ with sufficiently many sections, one can 
determine the second Chern class by choosing $r-2$ generic sections, and finding
the codimension-2 locus where those sections fail to be linearly independent.
Since twisting by $\calo(1)$ will alter the image in the Chow group only
trivially, we can always arrange for sufficiently many sections by twisting
with $\calo(n)$ for $n$ large enough. For bundles such as $E_\pm(n)$, we can
conveniently find sections\footnote{This was initially suggested to us by Nick 
Warner, and anticipated also by Duco van Straten.} by using the 2-periodicity 
of the complex \eqref{neq3} as the image of $Q$ in the previous step.

After some algebra, we find that the second Chern classes can be represented
as
\begin{equation}
c_2(E_+) - c_2(E_-) = [C_+-C_-] \in {\rm CH}^2(X_\psi) ,
\end{equation}
where $C_\pm$ stands for the algebraic curve
\begin{equation}
\eqlabel{cplusminus}
C_\pm = \{ x_1+x_2=0,x_3+x_4=0,x_5^2\pm\sqrt{5\psi} x_1x_3=0\}\subset X_\psi .
\end{equation}

Of course, we are really interested in the matrix factorizations and
corresponding bundles as objects in $D^b(Y)$, where $Y=X_\psi/(\zet_5)^3$ 
is the mirror quintic. Their second Chern classes take values in 
${\rm CH}^2(Y)$, and can be described by considering the image of $C_\pm$ 
under the $(\zet_5)^3$ orbifold group. We will study this quotient procedure 
carefully in the next section.

\section{Main Computation}
\label{main}

As before, we let $X_\psi$ be the one-parameter family of quintics given by 
\eqref{onepar}. The intersection of $X_\psi$ with the plane 
$P=\{x_1+x_2=x_3+x_4=0\}$ is a plane curve of degree $5$ which is reducible, 
with $3$ components (see left of Figure \ref{figure:Cplusminus}). One component 
is the line $x_5=0$, the other two are conics $C_\pm$ described by 
\eqref{cplusminus}. Obviously, $[C_+-C_-]=0 \in H_2(X_\psi)$ for all $\psi$, 
and thus the cycle $C_+-C_-$ defines a normal function, $\nu$ for the 
one-parameter family of quintics $X_\psi$. Consequently, we also obtain
a pair of curves and a normal function for the mirror quintic $Y$, which 
we will denote by the same symbols. Now pick a family of three-chains 
$\Gamma\subset Y$ with $\del\Gamma=C_+-C_-$. The domainwall tension or
truncated normal function is given by
\begin{equation}
\calt_B = \calt_B(z) = \int_{\Gamma} \hat\Omega ,
\end{equation}
where $\hat\Omega$ is a particular choice of holomorphic three-form on $Y$, 
further specified below.

The main result of our paper is that the Picard--Fuchs operator from
\eqref{pfop},
\begin{equation}
\eqlabel{PF}
\call = \theta^4-5z(5\theta+1)(5\theta+2)(5\theta+3)(5\theta+4)
\qquad\qquad \theta = zd/dz
\end{equation}
acting on $\calt_B(z)$ gives
\begin{equation}
\eqlabel{ePF}
\call \calt_B (z) = \frac{15}{16\pi^2} \sqrt{z}
\end{equation}
(as usual, $z=(5\psi)^{-5}$) where the constant is precisely the one 
in \eqref{inhomopf}. We conclude that $\calt_B(z)$ coincides with 
$\varpi_0(z)\calt_A(z)$, {\it up to a solution of the homogeneous 
Picard--Fuchs equation}. This is not unexpected since the choice of 
$\Gamma$ is ambiguous by $H^3(Y;\zet)$, so $\calt_B$ is ambiguous 
by an integral period. The claim that $\calt_B(z)-\varpi_0(z)\calt_A(z)$ 
is indeed an integral period will follow from the analytic continuation 
performed in \cite{open} and the boundary conditions on $\calt_B$ as we 
shall discuss below.

\subsection{Sketch of computation}
\label{sketch}

The strategy for proving \eqref{ePF} is to use the representation of
the holomorphic three-form on the hypersurface $\{W=0\}$ as 
the residue of a meromorphic 4-form $\tilde \Omega$ on projective space 
(Griffiths--Dwork method). The domainwall tension, which is defined by 
integrating the holomorphic three-form over a three-chain $\Gamma$ in 
$Y$ with $\del\Gamma=C_+-C_-$ can then be obtained by integrating 
$\tilde\Omega$ over a 4-chain which is a tube in $\projective^4\setminus 
\{W=0\}$ around $\Gamma$.\footnote{We are here temporarily confusing 
the mirror quintic with the family $X_\psi$.
The homogeneous Picard--Fuchs equation does not depend on this. The 
inhomogeneous term however does, see below.} By following the usual 
steps in the derivation of the Picard--Fuchs equation (see, \eg, 
\cite{guide}), the action of $\call$ on the domainwall tension can be 
reduced to a boundary term consisting of the integral of certain 
meromorphic three-forms over a tube around the boundary curves $C_\pm$. 
To be specific, let us consider the contribution from $C_+$. The main 
observation that will make the computation possible is the following. 

The curve $C_+$ lies in the plane $P=\{x_1+x_2=x_3+x_4=0\}$. Therefore,
if we could fit the tube around $C_+$ completely inside of $P$, 
the integral over it of any meromorphic three-form with poles on $W=0$ 
would vanish. The reason we {\it cannot} restrict the computation to 
$P$ is of course that $P\cap\{W=0\}$ contains not just $C_+$, but also 
$C_-$, as well as the line $x_5=0$, so that a tube around $C_+$ inside of $P$
will intersect one of the other components. But then, we can fit the tube
around $C_+$ into $P$ except for a small neighborhood of the points
where the components of $P\cap\{W=0\}$ meet. There are two such points,
$p_1=\{x_1=-x_2,x_3=x_4=x_5=0\}$ and $p_2=\{x_1=x_2=x_5=0,x_3=-x_4\}$,
and the computation can be localized to a small neighborhood of $p_1$
and $p_2$, which fit entirely inside an affine patch.

There is, however, an important subtlety in performing this computation
as we have just sketched.\footnote{We can attest to the fact that if this 
subtlety is ignored, a wrong answer is obtained!}  Namely, the intersection 
points $p_1$ and $p_2$ are actually singular points of the mirror quintic, 
and these singularities must be resolved first in order to perform the
computation. Recall that resolving the singularities amounts to varying 
the K\"ahler class on the quintic mirror to a generic value; since the 
inhomogeneous Picard--Fuchs equation should be independent of the K\"ahler 
class, it won't matter how we do the resolution of singularities.

\subsection{Resolution of singularities}
\label{res}

Since the plane $P=\{x_1+x_2=x_3+x_4=0\}$ itself plays an important r\^ole
in the computation, we also need to resolve singularities that appear on
it after passing to the quotient. The symmetry group $(\mathbb{Z}_5)^3$ 
permutes $25\cdot \frac{5!}{2!2!}=750$ similar planes, but a $\mathbb{Z}_5$ 
subgroup preserves our plane, with a generator acting via
\[ (x_1,-x_1,x_3,-x_3,x_5)\mapsto(x_1,-x_1,e^{2\pi i/5}x_3,-e^{2\pi i/5}x_3,
e^{-4\pi i/5}x_5).\]
This group action has three fixed points, at $p_1$, $p_2$, and $(0,0,0,0,1)$, 
and the first two of these must be resolved.\footnote{The third point does not 
lie on the quintic mirror for generic $\psi$, and need not be resolved.}

\begin{figure}[t]

\begin{center}
\includegraphics[width=2.8in]{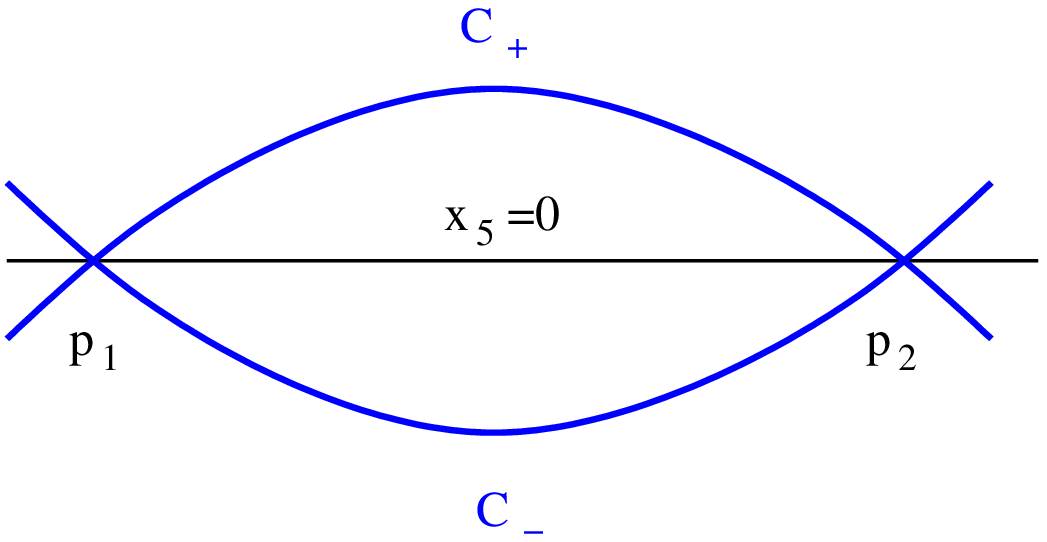}
\qquad
\includegraphics[width=2.8in]{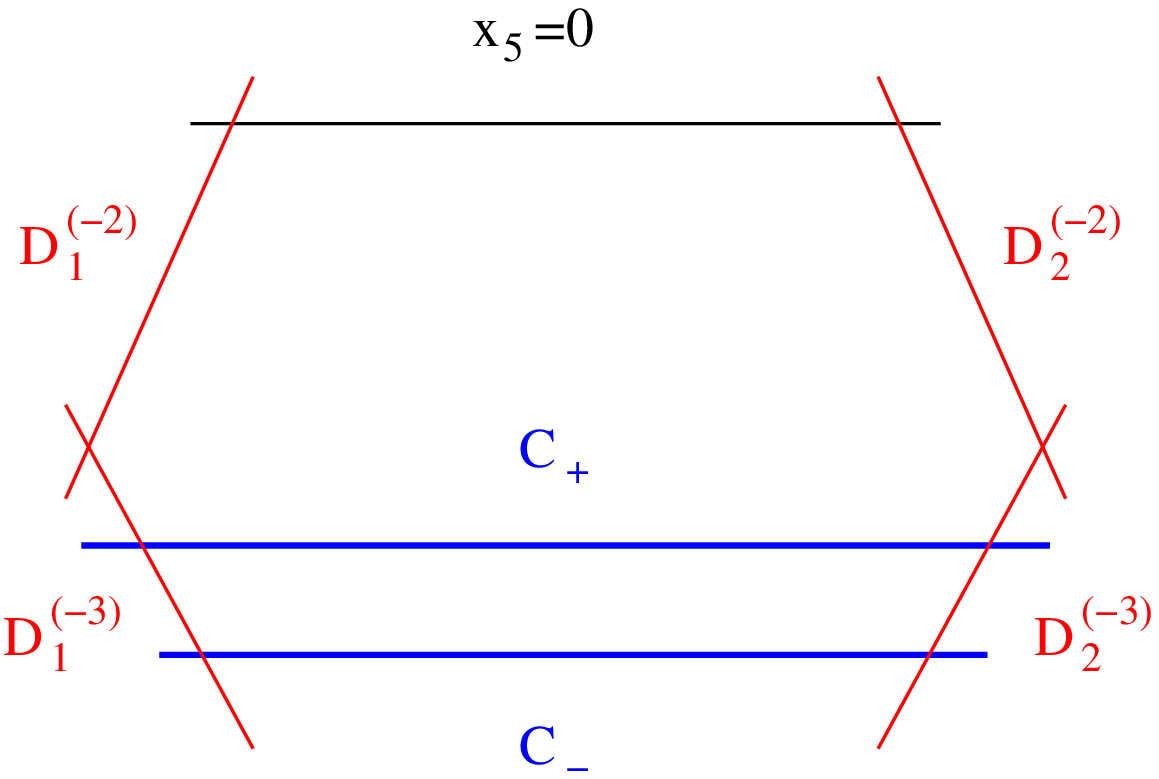}

\smallskip

\caption{The curves $C_\pm$ on $P=\complex\projective^2$ (quintic) and on 
$\widehat{S}=\widehat{\mathbb{CP}^2/\mathbb{Z}_5}$ (mirror quintic).}
\label{figure:Cplusminus}
\end{center}
\end{figure}

These singularities on $S=P/\zet_5$ are Hirzebruch--Jung singularities 
\cite{MR0062842,jung}, and can be resolved by classical methods\footnote{See 
\cite{Harvey:2001wm} for a recent discussion in the physics
literature.  In fact, the example in Figure~2 of \cite{Harvey:2001wm}
is exactly the case we must consider here.}  to obtain a surface
$\widehat S=\widehat{\mathbb{CP}^2/\mathbb{Z}_5}$. The result is that 
each singular point $p_i$ is replaced by two rational curves
$D_i^{(-2)}$ and $D_i^{(-3)}$, in the configuration shown in 
Figure~\ref{figure:Cplusminus}. We denote the intersections of (the
transforms of) $C_\pm$ with the curve $D_i^{(-3)}$ by $p_{i,\pm}$.

Resolving the quintic mirror itself is more involved, and we defer the
explicit computation to an appendix.  The result, however, is the
existence of two coordinate charts for the quintic mirror: the first
has coordinates
\begin{equation}
\eqlabel{patch1}
\begin{split} 
T &= x_1^{-1}x_2\\
X &= x_1x_3^{-2}x_4^3x_5^{-2}\\
Y &= x_1^{-5}x_5^5\\
Z &= x_1x_3^3x_4^{-2}x_5^{-2},
\end{split}
\end{equation}
and polynomial
\[ \frac15\left(1+T^5
+X^2 Y^2 Z^{3}
+X^{3} Y^2 Z^{2}
+ Y
\right)-\psi TXYZ.
\]
The resolution $\widehat{S}$ of the surface $S$ is given by $T=-1$,
$Z=-X$, and the restriction of the polynomial to $\widehat{S}$ is
\[ \frac15Y-\psi X^2Y=\frac15Y(1+\sqrt{5\psi}X)(1-\sqrt{5\psi}X).
\]
The points $p_{1,\pm}$ are given by $X=\pm\frac1{\sqrt{5\psi}}$, $Y=0$.
In the other coordinate chart, we have coordinates
\begin{equation}
\eqlabel{patch2}
\begin{split}
T' &= t' = x_3^{-1}x_4\\
X' &= (u')^{-2}(v')^{3}(w')^{-2} = x_1^{-2}x_2^3x_3x_5^{-2}\\
Y' &= (w')^{5} = x_3^{-5}x_5^5\\
Z' &= (u')^{3}(v')^{-2}(w')^{-2} = x_1^3x_2^{-2}x_3x_5^{-2},
\end{split}
\end{equation}
and polynomial
\[ \frac15\left(
(X')^2 (Y')^2 (Z')^{3}+(X')^{3} (Y')^2 (Z')^{2}
1+(T')^5+
+ Y'
\right)-\psi TXYZ.
\]
The resolution $\widehat{S}$ of the surface $S$ is given by $T'=-1$,
$Z'=-X'$, and the restriction of the polynomial to $\widehat{S}$ is
\[ \frac15Y'-\psi (X')^2Y'=\frac15Y'(1+\sqrt{5\psi}X')(1-\sqrt{5\psi}X').
\]
The points $p_{2,\pm}$ are given by $X'=\pm\frac1{\sqrt{5\psi}}$, $Y'=0$.

\subsection{Inhomogeneous Picard--Fuchs via Griffiths--Dwork}
\label{key}

Let's recall our conventions. We have
\begin{equation}
W = \frac{1}{5} \bigl(x_1^5+x_2^5+x_3^5+x_4^5+x_5^5\bigr) - \psi x_1x_2x_3x_4 x_5 ,
\end{equation}
and $z=(5\psi)^{-5}$. To derive the Picard--Fuchs equations by the Griffiths--Dwork
method, we introduce the four form on $\projective^4$,
\begin{equation}
\omega = \sum_{i} (-1)^{i-1} x_i dx_1\wedge\ldots\wedge\widehat{dx_i}\wedge\ldots 
\wedge dx_5 ,
\end{equation}
as well as the contraction of $\omega$ with the tangent vectors $\partial_i$ 
($i=1,\ldots,5$)
\begin{equation}
\omega_i = \omega(\partial_i) .
\end{equation}
A convenient choice of gauge for the holomorphic three-form is
\begin{equation}
\eqlabel{convenient}
\Omega (z) = {\rm Res}_{W=0} \tilde\Omega(z)\qquad\qquad
\text{where }
\tilde\Omega(z) := \frac{\omega}{W(z)} .
\end{equation}
Traditionally, one derives the Picard--Fuchs equation by working with the expression 
\eqref{convenient}, thought of as living on the quintic $X_\psi$. The holomorphic 
three-form on the mirror quintic $Y$ can be obtained by pulling back \eqref{convenient}
in local patches via blowup maps such as described in the appendix. For ordinary 
periods, the net effect of the quotient by $(\zet_5)^3$ is then simply an additional 
normalization factor of $5^{-3}$ \cite{cdgp}. Such a simple relation is not expected 
to hold for generic normal functions, so we need to evaluate things more carefully.

Following the reduction of pole algorithm of Griffiths, and keeping track of exact
pieces, we find with the above definitions,
\begin{equation}
\eqlabel{our}
\tilde\call \tilde\Omega := \bigl((1-\psi^5)\partial_\psi^4 - 10 \psi^4 \partial_\psi^3
-25\psi^3\partial_\psi^2 - 15\psi^2 \partial_\psi -1\bigr)\tilde\Omega 
= - d\tilde\beta ,
\end{equation}
where the exact piece is 
\begin{equation}
\eqlabel{exact}
\begin{split}
\tilde\beta = & \frac{3!}{W^4}\bigl(
x_2^4x_3^4x_4^4x_5^4\omega_1+\psi x_2x_3^5x_4^5x_5^5\omega_2 +
\psi^2 x_1x_2x_3^2x_4^6x_5^6\omega_3 \\&\qquad\qquad
+\psi^3 x_1^2x_2^2x_3^2x_4^3x_5^7\omega_4 + 
\psi^4 x_1^3x_2^3x_3^3x_4^3x_5^4\omega_5\bigl)\\
&+\frac{2}{W^3}\bigl(\psi x_3x_4^5x_5^5\omega_3 + 3\psi^2 x_1x_2x_3 x_4^2x_5^6\omega_4
+ 6\psi^3 x_1^2x_2^2x_3^2x_4^2x_5^3\omega_5\bigr)\\
&+\frac{1}{W^2}\bigl(\psi x_4x_5^5\omega_4 + 7\psi^2 x_1x_2x_3x_4x_5^2\omega_5\bigr) \\
&+ \frac{1}{W}\bigl(\psi x_5\omega_5\bigr) .
\end{split}
\end{equation}
Now the standard Picard--Fuchs operator $\call$ from \eqref{pfop} is related to $\tilde\call$
from \eqref{our} by
\begin{equation}
\eqlabel{comb1}
\call = -\frac{1}{5^4}\tilde\call\frac{1}{\psi} .
\end{equation}
On the other hand, the normalization of the holomorphic three-form in which the solutions
\eqref{fundamental} correspond to primitive integral periods of the mirror quintic is
\cite{cdgp}
\begin{equation}
\eqlabel{comb2}
\hat\Omega = \Bigl(\frac{5}{2\pi\ii}\Bigr)^3 \psi\, \Omega
= \Bigl(\frac{5}{2\pi\ii}\Bigr)^3 \psi\, {\rm Res}_{W=0}\frac{\omega}{W} .
\end{equation}
The domainwall tension for which we claim \eqref{ePF} is defined by
\begin{equation}
\calt_B(z) = \int_\Gamma\hat\Omega(z) ,
\end{equation}
where $\Gamma$ is any three-chain in $Y$ with $\del\Gamma=C_+-C_-$. Let $T_\epsilon(\Gamma)$
be a small tube around $\Gamma$ of size $\epsilon>0$. Then by \eqref{convenient},
\begin{equation}
\int_\Gamma\Omega = \frac{1}{2\pi\ii} \int_{T_\epsilon(\Gamma)}\tilde \Omega .
\end{equation}
By combining this with \eqref{comb1}, \eqref{comb2}, the claim \eqref{ePF} takes the form
\begin{equation}
\eqlabel{now}
\tilde\call\int_{T_\epsilon(\Gamma)}\tilde\Omega = -\frac{3\pi^2}{5^{1/2} \psi^{5/2}} ,
\end{equation}
which we now proceed to show.

There are two types of contributions to the RHS of \eqref{now}, depending on
whether the derivatives in $\tilde\call$ act on the chain or on $\tilde\Omega$. 
When $\tilde\call$ acts entirely on $\tilde\Omega$, we use \eqref{our} and obtain 
the boundary term
\begin{equation}
\eqlabel{boundary1}
- \int_{T_\epsilon(C_+-C_-)} \tilde\beta
\end{equation}
We will see below that this in fact gives the entire contribution claimed in \eqref{now}.
To show that the contributions from derivatives acting on $T_\epsilon(\Gamma)$ 
vanish, we use the fact that as $\psi$ varies, the three-chain $\Gamma$ changes to 
first order only at its boundary, in a way dictated by the dependence of $C_\pm$ on
$\psi$. Namely, the first order variation of $C_\pm$ is a section $n\in
N_{C_\pm/Y}$ of the normal bundle of $C_\pm$ in $Y$. This normal vector lifts
to the tube $T_\epsilon(C_\pm)$, and we shall show below that for $l=0,1,2,3$,
\begin{equation}
\eqlabel{boundary2}
\int_{T_\epsilon(C_+-C_-)} \frac{(x_1x_2x_3x_4x_5)^l \omega(n)}{W^{l+1}} = 0 ,
\end{equation}
where $\omega(n)$ is the contraction of $\omega$ with the normal vector $n$.
Establishing this claim together with the fact that \eqref{boundary1} evaluates 
to the RHS of \eqref{now} will complete the proof.

As described in subsection \ref{sketch}, we can evaluate integrals of
meromorphic three-forms over $T_\epsilon(C_\pm)$ as in \eqref{boundary1}
\eqref{boundary2}, by laying the tube into the plane $P$ (or rather its 
resolution $\widehat S$) outside a small neighborhood of the points
$p_{i,\pm}$. In those neighborhoods, we can use the coordinates of subsection
\ref{res}. Consider $p_{1,+}$, with coordinates \eqref{patch1}. The curve 
$C_+$ is given by $T=-1$, $X=-Z=1/\sqrt{5\psi}$ and locally parametrized by 
\begin{equation}
Y = r\ee^{\ii\varphi}
\end{equation}
varying in a neighborhood of $r=0$. Our tube $T_\epsilon(C_+)$ is defined
by picking a $C^\infty$ normal vector $v$ which satisfies $d_v W\neq 0$ on 
$C_+$ and points inside of $P$ outside of a small neighborhood of $Y=0$. 
To this end, let $f(r)$ be a non-negative $C^\infty$ function with $f(0)=1$ 
and $f(r)=0$ for $r\ge r_*>0$. We then choose 
\begin{equation}
\eqlabel{normal}
v = \frac{f(r)}{1+\frac{Y}{5}} \del_T -\ee^{-\ii\varphi}\del_X
+ \ee^{-\ii\varphi}\del_Z .
\end{equation}
Clearly, $v$ points inside of $P$ for $r>r_*$ and one easily checks
\def\sqrtpf{{\textstyle\sqrt{\frac{\psi}{5}}}}
\begin{equation}
d_v W|_{C_+} = f(r) + 2\sqrtpf\, r > 0
\qquad \text{for $0\le r \le 2 r_*$} .
\end{equation}
(We are here assuming that $\psi>0$. This is no restriction as long
as $\psi\neq 0$.) So the part of the tube $T_\epsilon(C_+;p_{1,+})$
around $C_+$ which is close to $p_{1,+}$ is parametrized as
\begin{equation}
\eqlabel{tube}
T=-1+\epsilon \ee^{\ii\chi} \frac{f(r)}{1+\frac Y5}\,,
\qquad X=-Z=\frac{1}{\sqrt{5\psi}}-\epsilon\ee^{\ii\chi}\ee^{-\ii\varphi} ,
\end{equation}
\begin{equation}
0\le \chi < 2\pi\,, \qquad
0\le \varphi < 2\pi\,, \qquad
0\le r < 2 r_* .
\end{equation}
(In all of this, we should really be taking the limit $\epsilon\to 0$, but 
the result will turn out to be independent of $\epsilon$.) There is then a 
corresponding piece of the tube around $p_{2,+}$. The part of the tube in 
between does not matter as it lies entirely within $P$, so any meromorphic 
three-form vanishes there. Finally, the contribution from $C_-$ will come 
from substituting $\sqrt{\psi}\to -\sqrt{\psi}$ in the final answer. 

We now apply the coordinate transformation \eqref{patch1} to evaluate
the three-forms $\omega_i$ on the tube \eqref{tube}. Choosing $x_1=1$,
we have
\begin{equation}
\begin{split}
\omega_1 &= -x_2dx_3dx_4dx_5+x_3dx_2dx_4dx_5-x_4dx_2dx_3dx_5+x_3dx_2dx_3dx_4 \\
\omega_2 &= dx_3dx_4dx_5\\
\omega_3 &= - dx_2dx_4dx_5\\
\omega_4 &=dx_2dx_3dx_5\\
\omega_5 &= -dx_2dx_3dx_4 ,
\end{split}
\end{equation}
and 
\begin{equation}
\begin{split}
\frac{dx_2}{x_2} & = \frac{dT}{T} \\
\frac{dx_3}{x_3} & = \frac{3}{5} \frac{dZ}{Z} + \frac{2}{5}
\frac{dX}{X} + \frac{2}{5} \frac{dY}{Y} \\
\frac{dx_4}{x_4} &= \frac{3}{5}\frac{dX}{X}+ \frac{2}{5}\frac{dZ}{Z}
+\frac{2}{5} \frac{dY}{Y} \\
\frac{dx_5}{x_5} &= \frac{1}{5}\frac{dY}{Y} .
\end{split}
\end{equation}
After restricting to $X=-Z$, this yields, $\omega_1=\omega_2=\omega_5=0$ and
\begin{equation}
\omega_3=\omega_4 = dx_2dx_3dx_5 = \frac{x_2x_3x_5}{5 TXY}\, dT dX dY .
\end{equation}
Substituting \eqref{tube}, we obtain
\begin{equation}
dTdXdY = \epsilon^2 \ee^{2\ii\chi} \frac{f}{1+\frac Y5}\Bigl(\frac{rf'}{f}-1\Bigr) 
d\chi d\varphi dr ,
\end{equation}
where $f'=df/dr$. The procedure to compute integrals of the forms
$p dx_2dx_2dx_5/W^{l+1}$, where $p$ is some monomial in $x_i$'s, over the 
tube $T_\epsilon(C_+;p_{1,+})$ is to first write a Laurent series in powers of 
$\epsilon\ee^{\ii\chi}$ and $\ee^{\ii\varphi}$. Integration over $\chi,\varphi$ 
will then retain only terms of order $\ee^{0\ii\chi}$, and $\ee^{0\ii\varphi}$, 
respectively. Finally, we'll do the integral over $r$.

To begin with, on the tube we have the expansion
\begin{equation}
\eqlabel{expan1}
W= \tilde\epsilon\bigl(f+2\sqrtpf\, r\bigr) - \tilde\epsilon^2
\bigr(2\tilde f^2+2\sqrtpf\,\tilde f r + \psi r \ee^{-\ii\varphi}\bigr)
+\tilde\epsilon^3\bigl(2\tilde f^3+\psi\tilde f r \ee^{-\ii\varphi}\bigr)
+\calo(\tilde\epsilon^4) ,
\end{equation}
where $\tilde\epsilon= \epsilon\ee^{\ii\chi}$ and $\tilde f=\frac{f}{1+\frac Y5}$.
In \eqref{expan1}, we have truncated to order $\epsilon^3$ since $\omega_3\propto 
\epsilon^2$, and the highest power of $W$ of interest corresponds to $l=3$. 

Let us consider the computation of a sample term in $\tilde\beta$ from
\eqref{exact}. Expanding in $\tilde\epsilon$, we have
\begin{equation}
\frac{\psi x_4 x_5^5\omega_3}{W^2} = 
\Bigl(
-\sqrtpf \ee^{\ii\varphi} r \frac{r f' -  f}{1+\frac Y5}
(f+2\sqrtpf r)^{-2}+\calo(\tilde\epsilon)\Bigr) d\chi d\varphi dr .
\end{equation}
The integration over $\varphi$ clearly kills this term. In fact, it
turns out that all the terms in $\tilde\beta$ which don't already vanish 
after restricting to $T_\epsilon(C_+;p_1)$ give zero after 
integration over $\chi$ and $\varphi$.

Going to $p_{2,+}$, where the local coordinates are \eqref{patch2} can be 
accomplished in the above formulas by exchanging $x_3$ with $x_1$ and $x_4$ 
with $x_2$. There are then only two terms to consider.

$\bullet$ The term $\frac{6 x_2^4x_3^4x_4^4x_5^4\omega_1}{W^4}$ gives, after 
integration over $\chi$ and $\varphi$,
\begin{equation}
(2\pi)^2 12 (rf'-f) r^2\left[\frac{\psi r^2 + 4\sqrt{5\psi} r f + 15 f^2}
{125\psi\bigl(f+2\sqrtpf r\bigr)^6}\right] .
\end{equation}
Integration over $r$ then gives
\begin{equation}
\eqlabel{yeah}
\frac{3\pi^2}{2\sqrt{5}\psi^{5/2}} .
\end{equation}

$\bullet$ The term $\frac{6\psi x_2 x_3^5x_4^5x_5^5\omega_2}{W^4}$ gives 
some complicated expression after integration over the angles, but the
integral over $r$ vanishes.

Taking into account the contribution from $C_-$, the final result 
for \eqref{boundary1} is
\begin{equation}
\eqlabel{together}
-\int_{T_\epsilon(C_+-C_-)} \tilde\beta = -\frac{3\pi^2}{\sqrt{5}\psi^{5/2}} ,
\end{equation}
precisely as claimed.

To show \eqref{boundary2}, we note that the normal vector implementing
first order deformation of $C_+$ is given by
\begin{equation}
n = -\frac{x_5^2}{\sqrt{5\psi}}\frac{1}{2\psi} \del_3 + 
\frac{x_5^2}{\sqrt{5\psi}}\frac{1}{2\psi}\del_4 .
\end{equation}
Thus, we find
\begin{equation}
\eqlabel{anyway}
\del_\psi^l\tilde\Omega(n) =
l!\frac{(x_1x_2x_3x_4x_5)^l}{W^{l+1}}\frac{x_5^2}{2\sqrt{5}\psi^{3/2}}
\bigl(\omega_3-\omega_4\bigr) .
\end{equation}
The expression \eqref{anyway} vanishes after restriction to the
tube, on which $\omega_3=\omega_4$ holds.

\subsection{Boundary conditions and monodromy}

We have just derived that the domainwall tension of the normal function 
associated with $C_+-C_-$ satisfies the same inhomogeneous Picard--Fuchs
equation \eqref{ePF} as the generating function for open Gromov--Witten 
invariants of the real quintic, \eqref{inhomopf}. This shows that
\begin{equation}
\eqlabel{showsthat}
\calt_B(z) = \varpi_0(z) \calt_A(t(z))
\end{equation}
{\it up to a solution of the homogeneous Picard--Fuchs equation}.
Identification of the normal function requires equality modulo
{\it periods}, which is a stronger statement. To establish it,
we need to determine a sufficient number of boundary conditions 
on $\calt_B(z)$. (The boundary conditions on $\calt_A$ are given
by \eqref{lvasym}.)

To fix this result, we make an explicit choice of three-chain
connecting $C_+$ and $C_-$. This is most easily done at $\psi=0$,
since $C_+$ and $C_-$ degenerate there (see \eqref{cplusminus}).
The Landau-Ginzburg monodromy $\psi\to \ee^{2\pi\ii/5}\psi$ 
interchanges $C_+$ with $C_-$. The natural choice of three-chain 
is therefore one that vanishes at $\psi = 0$, and changes orientation 
under the monodromy.

Now note that in our choice of gauge \eqref{comb2}, the solutions
of the Picard--Fuchs equation $\call\varpi=0$ actually all vanish 
as $\psi^k\sim z^{-k/5}$ for some $k=1,2,3,4$ as $\psi\to 0$. More 
precisely, the {\it integral periods}, known from \cite{cdgp}, 
vanish as $\psi^1\sim z^{-1/5}$, and are cyclically permuted by the 
Landau--Ginzburg monodromy $\psi\to\ee^{2\pi\ii/5}\psi$. We also
know however that the manifold itself is not singular at $\psi=0$,
so none of these vanishing periods corresponds to a vanishing cycle.
The integral over the three-chain should therefore vanish faster than
any period, and just change sign under the monodromy. The unique 
solution of \eqref{ePF} with these properties is given by
\begin{equation}
\calt_B(z) =\tau^{\rm orb}(z) = -\frac{4}{3}\sum_{m=0}^\infty 
\frac{\Gamma(-3/2-5m)}{\Gamma(-3/2)}
\frac{\Gamma(1/2)^5}{\Gamma(1/2-m)^5} z^{-(m+1/2)} .
\end{equation}
The explicit analytic continuation done in \cite{open} now shows
that $\tau^{\rm orb}(z)$ represents the same solution as 
$\omega_0(z)\calt_A(t(z))$, up to an integral period that
depends on the path chosen to connect $\psi=0$ with $\psi=\infty$.

\section{Summary and Conclusions}
\label{summary}

In this paper, we have explained why the superpotential/domainwall
tension for D-branes wrapped on compact Calabi--Yau manifolds will in
general satisfy a differential equation which is an extension
of the ordinary Picard--Fuchs equation. This relationship follows
from the insight that certain invariant holomorphic information 
about the topological D-brane boundary state, as a function of 
closed string moduli, is contained in the image of the algebraic
second Chern class under the Abel--Jacobi map to the intermediate
Jacobian, known Hodge theoretically as a normal function.
We have applied this formalism to the B-brane mirror to the real
quintic, and thereby re-derived the extended Picard--Fuchs equation
proposed in \cite{open}.

In combination with the proof of the enumerative predictions in the
A-model \cite{psw}, our results put open string mirror symmetry
for the real quintic \cite{open} at the same level as the
classical mirror theorems of Kontsevich, Givental, Lian-Liu-Yau,
and others. What is more, we have seen at several places very close 
connections to ideas from homological mirror symmetry. We have listed 
in section \ref{theory} several open problems that would make these
connections more concrete.

A somewhat unsatisfactory aspect of our derivation is that the
nature of the computation in section \ref{main} was severely
analytic. For many reasons, it would be desirable to develop 
a more algebraic understanding of extended Picard--Fuchs equations. 
The Griffiths infinitesimal invariant is likely to play an 
important role in such a development. Among other things, this 
might allow an easier generalization to other situations, especially 
if the expected connections with the categorical framework can be 
realized.

\begin{acknowledgments}
We would like to thank Pierre Deligne, Ezra Getzler, Phillip Griffiths, Jaya 
Iyer, Stefan M\"uller-Stach, Rahul Pandharipande, Tony Pantev, Duco van 
Straten, Richard Thomas, and Edward Witten for valuable discussions and 
communications. We are grateful to the Amsterdam Summer Workshop on String
Theory, the Simons Workshop in Mathematics
and Physics 2006, the KITP Santa Barbara during the program on String
Phenomenology, the Workshop on Homological Mirror Symmetry at IAS, the
Workshop on Enumerative Geometry at CRM in Montr\'eal, the 
Aspen Center for Physics, and the Simons Workshop in Mathematics and Physics 
2007, for providing a stimulating atmosphere during various stages of this 
project, and/or for the opportunity to present some preliminary versions 
of the results. The work of D.R.M.\ was supported in part by NSF
grant DMS-0606578. The work of J.W.\ was supported 
in part by the Roger Dashen Membership at the Institute for Advanced 
Study and by the NSF under grant number PHY-0503584.
Any opinions, findings, and conclusions or recommendations expressed in this 
material are those of the authors
and do not necessarily reflect the views of the National Science Foundation. 
\end{acknowledgments}

\section*{Appendix}
\addcontentsline{toc}{section}{Appendix}

In this appendix, we describe the resolution of singularities of the
quintic mirror, deriving the coordinate charts which are used in
making our key computation (see section \ref{key}).

The starting point is the singular model of the quintic mirror as
a hypersurface inside the singular ambient
space $\mathbb{CP}^4/(\mathbb{Z}_5)^3$.
Because the points $p_1$ and $p_2$ at which we wish to perform our
computation are singular points of this quotient, we need to carefully
resolve the singularities.  We will also explicitly resolve the
singularities on the surface $S=\mathbb{CP}^5/\mathbb{Z}_5$ defined by
$x_1+x_2=0$, $x_3+x_4=0$.

A consistent strategy for resolving singularities of the quintic mirror
was described in
Appendix B of \cite{guide}.  This strategy involves a choice of blowup,
and we will use the choice  described in \cite{geomaspects} rather than
that described in \cite{guide}.  

What makes the resolution tricky is that the ambient space
$\mathbb{CP}^4/(\mathbb{Z}_5)^3$ does not have a crepant resolution: the
coordinate vertices $(1,0,0,0,0)$ (and cyclic permutations) cannot
be resolved without introducing extraneous extra zeros into
the holomorphic form of top degree.  
However, the quintic mirror does not pass through those
points, so this fact does not prevent us from resolving the quintic
mirror itself.

Each of the points $p_1$ and $p_2$ lies in the fixed locus of a particular
$(\mathbb{Z}_5)^2$ subgroup of $(\mathbb{Z}_5)^3$.  Thus, we will describe
a coordinate chart on the blowup for each by describing the blowup
of the quotient by the $(\mathbb{Z}_5)^2$ subgroup, and indicating how
the quotient by the remaining $\mathbb{Z}_5$ is to be performed.

The point $p_1=(1,-1,0,0,0)$ is
contained in the affine chart $x_1=1$, and its stabilizer
is the $(\mathbb{Z}_5)^2$ subgroup of $(\mathbb{Z}_5)^3$
which fixes the affine coordinate $x_2/x_1$.

That is, we begin with affine coordinates
$t=x_2/x_1$, $u=x_3/x_1$, $v=x_4/x_1$, 
$w=x_5/x_1$ and the $(\mathbb{Z}_5)^2$ action on $(u,v,w)$ which preserves
the product $uvw$.  The rational functions invariants under
 this action are generated by
$t$, $u^5$, $v^5$, 
and $uvw$; the remaining $\mathbb{Z}_5$ in our full $(\mathbb{Z}_5)^3$ symmetry
group then preserves $u^5$ and $v^5$ while acting 
oppositely on $t$ and on $uvw$, so that the invariants under the
full group would include $t^5$ and $tuvw$.
 The polynomial defining 
the quintic mirror in this affine  coordinate chart is
\[ \frac15\left(1 + t^5 + u^5 + v^5 + w^5\right) - \psi tuvw ,\]
and our surface $S$ is defined by $t=-1$, $v=-u$.

\begin{figure}[h]

\begin{center}
\smallskip

\includegraphics[width=2.5in]{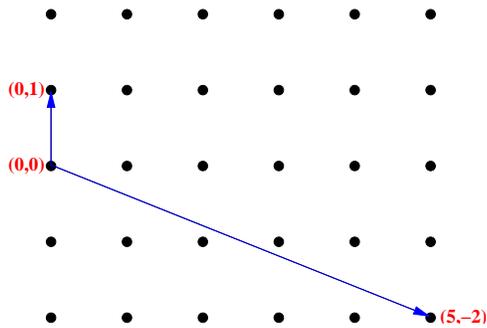}

\smallskip

\caption{
Toric data for the resolution of the $\mathbb{C}^2/\mathbb{Z}_5$ 
quotient (from \cite{Harvey:2001wm}).}%
\label{figure:hkmm}
\end{center}
\end{figure}

The group action on the surface $S$ is generated by
\[ (u,w)\mapsto (e^{2\pi i/5}u,e^{-4\pi {\it i}/5}w),\]
and the invariant rational monomials for this action are generated by
$w^5$ and $uw^{-2}$.  To describe the corresponding toric geometry,
we represent an arbitrary invariant rational monomial in the form
\[ (w^5)^a(uw^{-2})^b=u^bw^{5a-2b},\]
and note that the condition for this monomial to be regular, i.e.,
to have no pole at the origin, is
$b\ge0$, $5a-2b\ge0$.  These inequalities determine the toric data:
the dual vectors $(0,1)$ and $(5,-2)$ generate a cone consisting of
all inequalities satisfied by regular monomials, as depicted 
in Figure~\ref{figure:hkmm} (which was borrowed from \cite{Harvey:2001wm}).

To resolve the singularity, we subdivide this cone using cones whose
edges form a basis for the lattice $\mathbb{Z}^2$.  This can always
be done by using lattice elements which are close to the origin: in
our example, the three subcones are generated by (i) $(0,1)$ and $(1,0)$,
(ii) $(1,0)$ and $(3,-1)$, and (iii) $(3,-1)$ and $(5,-2)$.

The coordinates on these three charts have the property that the
inequalities defining which monomials are regular within the chart
are precisely spanned by the generators of the cone.  In our example,
the first chart has coordinates
\[ u_{\text{(i)}}=uw^{-2}, \quad w_{\text{(i)}}=w^5 \]
so that
\[ u^b w^{5a-2b} = (u_{\text{(i)}})^b (w_{\text{(i)}})^a;\]
the second chart has coordinates
\[ u_{\text{(ii)}}=u^3w^{-1}, \quad w_{\text{(ii)}}=w^2u^{-1}\]
so that
\[ u^bw^{5a-2b} = (u_{\text{(ii)}})^{a}(w_{\text{(ii)}})^{3a-b};\]
and the third chart has coordinates
\[ u_{\text{(iii)}}=u^5, \quad w_{\text{(iii)}}=wu^{-3}\]
so that
\[ u^bw^{5a-2b} =(u_{\text{(iii)}})^{3a-b}(w_{\text{(iii)}})^{5a-2b}.\]

The exceptional curve $D_1^{(-3)}$ is represented by 
$w_{\text{(i)}}=u_{\text{(ii)}}=0$, and has
self-intersection $-3$ due to the change of coordinates map
\[ u_{\text{(ii)}} = u_{\text{(i)}}^3 w_{\text{(i)}},
\quad w_{\text{(ii)}} = u_{\text{(i)}}^{-1}
.\]

The exceptional curve $D_1^{(-2)}$ is represented by 
$w_{\text{(ii)}}=u_{\text{(iii)}}=0$,  and has
self-intersection $-2$ due to the change of coordinates map
\[ u_{\text{(iii)}} = u_{\text{(ii)}}^2 w_{\text{(ii)}}, 
\quad w_{\text{(iii)}} = u_{\text{(ii)}}^{-1} 
.\]

The defining polynomial for the quintic mirror, when restricted to $S$,
takes the following form in these coordinate charts:
\begin{align*}
\frac15w_{\text{(i)}}-\psi u_{\text{(i)}}^2w_{\text{(i)}}
&=\frac15w_{\text{(i)}}(1-\sqrt{5\psi}u_{\text{(i)}})(1+\sqrt{5\psi}u_{\text{(i)}}),\\
\frac15 u_{\text{(ii)}}w_{\text{(ii)}}^3 -\psi u_{\text{(ii)}}w_{\text{(ii)}}
&=\frac15u_{\text{(ii)}}w_{\text{(ii)}}(w_{\text{(ii)}}-\sqrt{5\psi})(w_{\text{(ii)}}+\sqrt{5\psi}),\\
\frac15u_{\text{(iii)}}^3w_{\text{(iii)}}^5-\psi u_{\text{(iii)}}w_{\text{(iii)}}
&=\frac15u_{\text{(iii)}}w_{\text{(iii)}}(u_{\text{(iii)}}w_{\text{(iii)}}^2-\sqrt{5\psi})(u_{\text{(iii)}}w_{\text{(iii)}}^2+\sqrt{5\psi}).
\end{align*}
Thus, the intersection points $p_{1,\pm}$
of $C_\pm$ with $D_1^{(-3)}$ can be found
in either chart (i) at $(\pm (5\psi)^{-1/2},0)$ or chart (ii) at
$(0,\pm(5\psi)^{1/2})$.
All of this agrees with the illustration in Figure~\ref{figure:Cplusminus}.

We now turn to the resolution of the quintic mirror itself.
In order to describe the $\mathbb{C}^3/(\mathbb{Z}_5)^2$ 
quotient singularity in terms
of toric geometry, we represent an arbitrary invariant rational monomial in
the form
\[ (u^5)^a (v^5)^b (uvw)^c = u^{5a+c} v^{5b+c} w^c,\]
and note that the condition for this monomial to be regular
is $5a+c\ge0$, $5b+c\ge0$, and $c\ge0$.  Those three inequalities
determine the toric data: one takes the dual vectors $(5,0,1)$,
$(0,5,1)$, $(0,0,1)$ to these inequalities, 
and notes that all inequalities satisfied on the regular
functions are nonnegative linear combinations of these vectors.

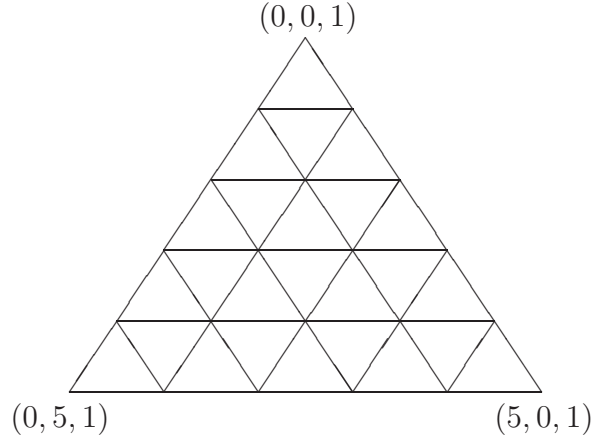
\begin{figure}[t]
\setlength{\unitlength}{.38em}

\begin{center}
\begin{picture}(40,30)

\thinlines

\put(-5,-3){$(0,5,1)$}
\put(36,-3){$(5,0,1)$}
\put(16,31){$(0,0,1)$}

\put(0,0){\line(1,0){40}}
\put(40,0){\line(-2,3){20}}
\put(20,30){\line(-2,-3){20}}

\put(4,6){\line(1,0){32}}
\put(32,0){\line(-2,3){16}}
\put(24,24){\line(-2,-3){16}}

\put(8,0){\line(-2,3){4}}
\put(32,0){\line(2,3){4}}
\put(16,24){\line(1,0){8}}

\put(16,0){\line(2,3){12}}

\put(16,0){\line(-2,3){4}}
\put(24,0){\line(2,3){4}}

\put(32,12){\line(-1,0){24}}

\put(32,12){\line(-2,-3){4}}
\put(20,18){\line(1,0){8}}

\put(12,18){\line(2,-3){12}}

\put(12,18){\line(1,0){8}}
\put(8,12){\line(2,-3){4}}

\end{picture}
\end{center}

\caption{Toric data for the resolution of the $\mathbb{C}^3/(\mathbb{Z}_5)^2$ 
quotient (from \cite{geomaspects}).}%
                         \label{figure:toric}

\end{figure}

The resolutions of toric geometry are obtained by subdividing the cone
generated by those vectors into cones whose generating vectors give a
basis for the lattice $\mathbb{Z}^3$.  There are a number of ways of
doing this, but we use the symmetric one illustrated in 
Figure~\ref{figure:toric} (which is borrowed from \cite{geomaspects}).  
The three dual vectors
$(5,0,1)$,
$(0,5,1)$, $(0,0,1)$
are the vertices of the large triangle,
and the resolution has coordinate charts determined by the small
triangles in the diagram.

There are two kinds of coordinate charts.  The first type of chart
$U_{\alpha\beta}$, labeled by
$\alpha$ and $\beta$ with $\alpha\ge0$, $\beta\ge0$ and
$\alpha+\beta\le4$, corresponds to the upward pointing
triangle in Figure~\ref{figure:toric} with vertices
$(\alpha,\beta+1,1)$, $(\alpha,\beta,1)$ and $(\alpha+1,\beta,1)$.
This chart will have coordinates $X_{\alpha\beta}$, $Y_{\alpha\beta}$,
$Z_{\alpha\beta}$, and $T=t$ satisfying
\[ 
(X_{\alpha\beta})^{a\alpha+b(\beta+1)+c}
(Y_{\alpha\beta})^{a\alpha+b\beta+c}
(Z_{\alpha\beta})^{a(\alpha+1)+b\beta+c}
= (u^5)^a(v^5)^b(uvw)^c;
\]
this can be solved for the coordinates, giving
\begin{align*}
T&=t\\
X_{\alpha\beta} &= u^{-\beta}v^{5-\beta}w^{-\beta}\\
Y_{\alpha\beta} &= u^{\alpha+\beta-4}v^{\alpha+\beta-4}w^{\alpha+\beta+1}\\
Z_{\alpha\beta} &= u^{5-\alpha}v^{-\alpha}w^{-\alpha}.
\end{align*}
The defining polynomial of the quintic mirror in this chart is
\[ \frac15\left(1+T^5
+X_{\alpha\beta}^\alpha Y_{\alpha\beta}^\alpha Z_{\alpha\beta}^{\alpha+1}
+X_{\alpha\beta}^{\beta+1} Y_{\alpha\beta}^\beta Z_{\alpha\beta}^{\beta}
+X_{\alpha\beta}^{4-\alpha-\beta} Y_{\alpha\beta}^{5-\alpha-\beta} Z_{\alpha\beta}^{4-\alpha-\beta}
\right)-\psi TX_{\alpha\beta}Y_{\alpha\beta}Z_{\alpha\beta}.
\]

The second type of chart
$\widetilde{U}_{\alpha\beta}$, labeled by
$\alpha$ and $\beta$ with $\alpha\ge0$, $\beta\ge0$ and $\alpha+\beta\le3$,
corresponds to the downward pointing triangles in Figure~\ref{figure:toric}
with vertices $(\alpha,\beta+1,1)$, $(\alpha+1,\beta+1,1)$ and 
$(\alpha+1,\beta,1)$.  
This chart will have coordinates $\widetilde{X}_{\alpha\beta}$, 
$\widetilde{Y}_{\alpha\beta}$,
$\widetilde{Z}_{\alpha\beta}$, and $\widetilde{T}=t$ satisfying
\[ 
(\widetilde{X}_{\alpha\beta})^{a\alpha+b(\beta+1)+c}
(\widetilde{Y}_{\alpha\beta})^{a(\alpha+1)+b(\beta+1)+c}
(\widetilde{Z}_{\alpha\beta})^{a(\alpha+1)+b\beta+c}
= (u^5)^a(v^5)^b(uvw)^c;
\]
this can be solved for the coordinates, giving
\begin{align*}
\widetilde{T}&=t\\
\widetilde{X}_{\alpha\beta} &= u^{\alpha-4}v^{\alpha+1}w^{\alpha+1}\\
\widetilde{Y}_{\alpha\beta} &= u^{4-\alpha-\beta}v^{4-\alpha-\beta}w^{-1-\alpha-\beta}\\
\widetilde{Z}_{\alpha\beta} &= u^{\beta+1}v^{\beta-4}w^{\beta+1}.
\end{align*}
The defining polynomial of the quintic mirror in this chart is
\[ \frac15\left(1+\widetilde{T}^5
+\widetilde{X}_{\alpha\beta}^\alpha \widetilde{Y}_{\alpha\beta}^{\alpha+1} \widetilde{Z}_{\alpha\beta}^{\alpha+1}
+\widetilde{X}_{\alpha\beta}^{\beta+1} \widetilde{Y}_{\alpha\beta}^{\beta+1} \widetilde{Z}_{\alpha\beta}^{\beta}
+\widetilde{X}_{\alpha\beta}^{4-\alpha-\beta} \widetilde{Y}_{\alpha\beta}^{3-\alpha-\beta} \widetilde{Z}_{\alpha\beta}^{4-\alpha-\beta}
\right)-\psi \widetilde{T}\widetilde{X}_{\alpha\beta}\widetilde{Y}_{\alpha\beta}\widetilde{Z}_{\alpha\beta}.
\]

To determine which chart we should use, we restrict the coordinates
on $U_{\alpha\beta}$ and $\widetilde{U}_{\alpha\beta}$ to the blowup
of $S$,
and express them as functions of the coordinates $u_{(i)}$, $w_{(i)}$
in the first coordinate chart of that blowup.
In $U_{\alpha\beta}$ we find
\begin{align*}
T&=1\\
X_{\alpha\beta} &= (-1)^{5-\beta}u_{(i)}^{5-2\beta}w_{(i)}^{2-\beta}\\
Y_{\alpha\beta} &= (-1)^{\alpha+\beta-4}u_{(i)}^{2\alpha+2\beta-8}w_{(i)}^{\alpha+\beta-3}\\
Z_{\alpha\beta} &= (-1)^{\alpha}u_{(i)}^{5-2\alpha}w_{(i)}^{2-\alpha},
\end{align*}
and from this we conclude that $U_{22}$ restricts to this coordinate
chart on $S$, and contains the points $p_{1,\pm}$.  
In $\widetilde{U}_{\alpha\beta}$ we find
\begin{align*}
\widetilde{T}&=1\\
\widetilde{X}_{\alpha\beta} &= (-1)^{\alpha+1}u_{(i)}^{2\alpha-3}w_{(i)}^{\alpha-1}\\
\widetilde{Y}_{\alpha\beta} &= (-1)^{4-\alpha-\beta}u_{(i)}^{8-2\alpha-2\beta}w_{(i)}^{3-\alpha-\beta}\\
\widetilde{Z}_{\alpha\beta} &= (-1)^{\beta-4}u_{(i)}^{2\beta-3}w_{(i)}^{\beta-1},
\end{align*}
and so none of the coordinate charts $\widetilde{U}_{\alpha\beta}$ is
a neighborhood of $p_{1,\pm}$.

Thus, the first coordinate chart we use will have coordinates
\begin{align*} 
T &= T_{22} = t = x_1^{-1}x_2\\
X &= X_{22} = u^{-2}v^{3}w^{-2} = x_1x_3^{-2}x_4^3x_5^{-2}\\
Y &= Y_{22} = w^{5} = x_1^{-5}x_5^5\\
Z &= Z_{22} = u^{3}v^{-2}w^{-2} = x_1x_3^3x_4^{-2}x_5^{-2},
\end{align*}
and polynomial
\[ \frac15\left(1+T^5
+X^2 Y^2 Z^{3}
+X^{3} Y^2 Z^{2}
+ Y
\right)-\psi TXYZ.
\]
The resolution $\widehat{S}$ of the surface $S$ is given by $T=-1$,
$Z=-X$, and the restriction of the polynomial to $\widehat{S}$ is
\[ \frac15Y-\psi X^2Y=\frac15Y(1+\sqrt{5\psi}X)(1-\sqrt{5\psi}X).
\]
The points $p_{1,\pm}$ are given by $X=\pm\frac1{\sqrt{5\psi}}$, $Y=0$.

To find a neighborhood of the points $p_{2,\pm}$, we use the affine chart
$x_3=1$, and label the coordinates on this chart as
$u'=x_1/x_3$, $v'=x_2/x_3$, $t'=x_4/x_3$, $w'=x_5/x_3$. The defining
polynomial becomes
\[ \frac15\left((u')^5+(v')^5+1+(t')^5+(w')^5
\right) -\psi t'u'v'w',
\]
and the surface $S$ is defined by $t'=-1$, $v'=-u'$.  We've chosen
the notation so that the $(\mathbb{Z}_5)^2$ subgroup of 
$(\mathbb{Z}_5)^3$ which stabilizes $p_2$ acts exactly as in the
previous case: acting on $(u',v',w')$ and preserving the product
$u'v'w'$.

Since the combinatorics are identical, the computation produces the
same result as in the first part of this appendix.  We find a neighborhood
of $p_{2,\pm}$ with coordinates
\begin{align*} 
T' &= t' = x_3^{-1}x_4\\
X' &= (u')^{-2}(v')^{3}(w')^{-2} = x_1^{-2}x_2^3x_3x_5^{-2}\\
Y' &= (w')^{5} = x_3^{-5}x_5^5\\
Z' &= (u')^{3}(v')^{-2}(w')^{-2} = x_1^3x_2^{-2}x_3x_5^{-2},
\end{align*}
and polynomial
\[ \frac15\left(
(X')^2 (Y')^2 (Z')^{3}+(X')^{3} (Y')^2 (Z')^{2}
1+(T')^5+
+ Y'
\right)-\psi TXYZ.
\]
The resolution $\widehat{S}$ of the surface $S$ is given by $T'=-1$,
$Z'=-X'$, and the restriction of the polynomial to $\widehat{S}$ is
\[ \frac15Y'-\psi (X')^2Y'=\frac15Y'(1+\sqrt{5\psi}X')(1-\sqrt{5\psi}X').
\]
The points $p_{2,\pm}$ are given by $X'=\pm\frac1{\sqrt{5\psi}}$, $Y'=0$.

Just for completeness, we include the change of coordinates between these
two charts.
\begin{align*}
T' &= \sqrt[5]{X/Z}\\
X' &= T^3 \sqrt[5]{X^2Z^3}\\
Y' &= 1/(X^2YZ^3)\\
Z' &= T^{-2} \sqrt[5]{X^2Z^3}.
\end{align*}
These formulas illustrate the important point that the additional
$\mathbb{Z}_5$ quotient must be considered in each case.

\newcommand{\refcno}[1]{}
\newcommand{\hyp}{-}
\colorlinksblue
\addcontentsline{toc}{section}{References}

\small

%\bibliographystyle{amsunsrt-ensc}
%\bibliography{normal}
 
\ifx\undefined\bysame
\newcommand{\bysame}{\leavevmode\hbox to3em{\hrulefill}\,}
\fi

\end{document}